\documentclass[a4paper,fleqn,usenatbib]{mnras}

\usepackage{newtxtext,newtxmath}

\usepackage[T1]{fontenc}
\usepackage{ae,aecompl}

\usepackage{graphicx}	
\usepackage{amsmath}	
\usepackage{amssymb}	
\usepackage{multirow}
\usepackage{listings}

\title[On the Oosterhoff dichotomy in the Galactic bulge]{On the Oosterhoff dichotomy in the Galactic bulge: I. spatial distribution}

\author[Prudil et al.]{
Z. Prudil$^{1}$\thanks{E-mail: prudilz@ari.uni-heidelberg.de}, I. D\'ek\'any$^{1}$, M. Catelan$^{2,3}$\thanks{On sabbatical leave at European Southern Observatory, Av. Alonso de C\'ordova 3107, 7630355 Vitacura, Santiago, Chile.}, R. Smolec$^{4}$, E. K. Grebel$^{1}$, M. Skarka$^{5,6}$ \\
$^{1}$ Astronomisches Rechen-Institut, Zentrum f{\"u}r Astronomie der Universit{\"a}t Heidelberg, M{\"o}nchhofstr. 12-14, D-69120 Heidelberg, Germany\\
$^{2}$ Instituto de Astrof{\'i}sica, Pontificia Universidad Cat{\'o}lica de Chile, Av. Vicu{\~n}a Mackenna 4860, 782-0436 Macul, Santiago, Chile\\
$^{3}$ Instituto Milenio de Astrof{\'i}sica, Santiago, Chile\\
$^{4}$ Nicolaus Copernicus Astronomical Center, Polish Academy of Sciences, ul. Bartycka 18, 00-716 Warszawa, Poland\\
$^{5}$ Department of Theoretical Physics and Astrophysics, Masaryk University, Kotl\'{a}\v{r}sk\'{a} 2, CZ-311 37, Czech Republic.\\
$^{6}$ Astronomical Institute, Czech Academy of Sciences, Fri\v{c}ova 298, CZ-251 65, Ond\v{r}ejov, Czech Republic\\
}

\date{Accepted XXX. Received YYY; in original form ZZZ}

\pubyear{2017}

\begin{document}

\label{firstpage}
\pagerange{\pageref{firstpage}--\pageref{lastpage}}
\maketitle

\begin{abstract}
We present a study of the Oosterhoff (Oo) dichotomy in the Galactic bulge using 8\,141 fundamental mode RR~Lyrae stars. We used public photometric data from the Optical Gravitational Lensing Experiment (OGLE) and the Vista Variables in the V\'ia L\'actea survey (VVV). We carefully selected fundamental mode stars without modulation and without association with any globular cluster located toward the Galactic bulge. Subsequently, we identified and separated the Oosterhoff groups I and II on the basis of their period-amplitude distribution and using a relation fitted to the Oosterhoff I locus. Both Oosterhoff groups were then compared to observations of two bulge globular clusters and with models of stellar pulsation and evolution. We found that some of the variables classified as Oo\,II belong to a third Oo group. The Oosterhoff II variables are more metal-poor on average, more massive, and cooler than their Oosterhoff I counterparts. The analysis of their spatial distribution shows a systematic difference between \textit{foreground}, central and \textit{background} regions in the occurrence of the Oosterhoff II group. The difference between the Oo\,I and II groups is also seen in their distance distributions with respect to the Galactic bar, but neither group is associated with the bar.
\end{abstract}




\section{Introduction}

The RR~Lyrae stars are helium-core burning, radially pulsating variables, found in old stellar systems. Thanks to their visual and infrared period-metallicity-luminosity relations, they are highly useful distance indicators within the Local Group of galaxies \citep{Alcock1997,Clementini2003,Monelli2017}. Their contribution has been recognized in studies involving the Milky Way and its components like the Galactic bulge \citep{Dekany2013,Pietrukowicz2015}, Galactic halo \citep{Sesar2013}, and Galactic disk \citep{Dekany2018}, and also the Large and Small Magellanic Clouds \citep{Haschke-2012-SMC-LMC,JD2massive016RRLyr}. Moreover, they have been proven to be useful in estimating the metallicity distribution function of a given system. Using their pulsation periods and light curve parameters we can estimate their individual metallicities \citep{JK1996,Smolec2005,Hajdu2018}. The RR~Lyrae stars are divided into three sub-classes based on their mode of pulsation: The most common fundamental-mode (RRab) variables, the first-overtone (RRc) variables, and the double-mode (RRd) pulsators, which pulsate simultaneously in the fundamental mode and the first overtone.

Although RR~Lyrae stars have been extensively studied in the past decades, there are still several open questions regarding their physical nature. For example, the uncertainty in their mass \citep[due to the lack of a dynamical mass determination in binary systems, see][]{Hajdu2015,Liska2016}, the modulation of their light curves known as the Blazhko effect \citep[see][for extensive reviews]{Szabo2014,Smolec2016BL}, and the Oosterhoff dichotomy \citep{Oosterhoff1939}. 

\citet{Oosterhoff1939} noticed that five Milky Way globular clusters (from hereon referred as GCs) can be divided into two classes based on the mean of the pulsation periods of RR~Lyrae stars within them. Later on, Oosterhoff extended his analysis using additional globular clusters with RR~Lyrae stars \citep{Oosterhoff1944}. Nowadays, many of the Milky Way globular clusters can be separated into objects containing either RR~Lyrae stars belonging to the Oosterhoff type I (Oo\,I) or Oosterhoff type II (Oo\,II) group. The Oo\,I clusters are more metal-rich ($\text{[Fe/H]}\geqslant-1.7$\,dex), contain RR~Lyraes that have short pulsation periods (around 0.55\,days for RRab stars), and seem to be deficient in first-overtone pulsators (which provide less than 20\,\% of the RR~Lyrae population). On the contrary, the Oo\,II clusters are mostly metal-poor ($\text{[Fe/H]}\leqslant-1.7$\,dex), they have RR~Lyraes with longer pulsation periods on average (0.65\,days for RRab stars) and a rather high fraction of RRc stars (around 40\,\%). 

Over the past decades several explanations of the Oosterhoff dichotomy have been proposed, for example, the existence of a hysteresis zone \citep{vanAlbada1973} implying two crossings of RR~Lyraes through the instability strip (IS). \citeauthor{vanAlbada1973} suggest that Oo\,II RR~Lyraes evolve from the blue to the red edge of the IS, while Oo\,I RR~Lyraes evolve in the opposite direction, from the red to the blue edge of the IS. In other studies, \citet{Sandage-K-S-1981} and \citet{Sandage1981} examined zero-age-horizontal branch models and suggested a higher helium abundance in Oo\,II clusters in order to explain the difference. Overall, a general consensus has not yet been reached \citep{Catelan2009}.

The aforementioned solutions are in tension with the observed properties of RR~Lyraes in the two GCs NGC~6441 and NGC~6388. These two GCs stand out from the classical Oosterhoff scheme since they contain RR~Lyrae stars with long pulsation periods and high metallicities, $\text{[Fe/H]}=-0.53\pm 0.11$\,dex and $\text{[Fe/H]}=-0.60\pm 0.15$\,dex for NGC~6441 and NGC~6388 \citep{ArmZin1988}, respectively. They create their own Oo\,III group \citep{Pritzl2000}. Interestingly, when we compare the population of RR~Lyraes in low-mass galaxies outside the Milky Way in the Local Group, we find that these objects do not fall into either Oosterhoff group. The Magellanic Clouds, other dwarf galaxies and their GCs fall between the classical Oosterhoff groups and create a so-called Oosterhoff-intermediate population \citep{Catelan2009}. Therefore, we no longer see a bimodality but rather an Oosterhoff continuum \citep[see left panel of fig.~6 in][]{Catelan2009}. This suggests that the Oo dichotomy is intrinsic to Galactic GCs and thus might be a unique fingerprint of the Milky Way's evolution history. 

In this paper, we explore the spatial structure of the Galactic bulge using RR~Lyrae stars with respect to their Oosterhoff association. Galactic bulges are commonly grouped into two main types, classical (spheroidal) bulge formed through early merging events and (bar-like) pseudobulges formed through buckling disk instabilities \citep{Kormendy2004}. The Milky Way's bulge has a boxy/peanut (B/P) shape and is mainly characterized by cylindrical rotation \citep{Howard2009}. Roughly half of the edge-on disk galaxies contain B/P bulges \citep{Lutticke2000}. The B/P shape of our Galaxy is directly visible due to the fact that the structure is viewed under a non-zero angle \citep[27\,deg,][]{Wegg2013}, and can be traced using red clump giants \citep{McWilliam2010,Gonzalez2011,Ness2012,Wegg2013}. Attempts to trace bar structure using old-population stars has been made in the past with disparate results \citep{Dekany2013,Pietrukowicz2015}. The radial velocity distribution of bulge RR Lyrae stars is inconsistent with typical B/P bulge kinematics, and overall they do not contribute to the boxy shape of the Galactic bulge \citep{Kunder2016}.

The Galactic bulge has been a target for studies of the Oosterhoff dichotomy in the past, for example, by \citet{Kunder2009}. Using $V$-band data from the MACHO survey for RRab stars with \textit{normal} light curves \citep[based on the calculated deviation parameter from][]{JK1996}, \citeauthor{Kunder2009} found that the difference in pulsation periods between the bulge Oo\,I variables and the globular cluster OoI variables is about $\Delta \text{log}P=0.02$\,days. In addition, in their study variables identified as Oo\,II stars could be on average brighter by 0.2\,mag than their Oo\,I counterparts. 

In this first paper of the series, we used data of RRab variables from the Optical Gravitational Lensing Experiment IV \citep[from hereon OGLE-IV,][]{Udalski2015} and the VISTA Variables in the V\'ia L\'actea survey \citep[VVV,][]{Minniti2010} to study the physical and spatial properties of the Oosterhoff populations in the Galactic bulge. In the second paper of the series, Paper\,II (Prudil et al., in preparation), we will look at the kinematical distribution of some of the studied stars from the present paper. This paper is structured as follows. In Section \ref{sec:SamSel} we describe how we selected our stellar sample. Section \ref{sec:DistOogr} discusses the computation of individual distances to the selected stars and the separation of the two Oosterhoff groups in the Galactic bulge. In Sec.~\ref{NGC6441-PhysiParam} we compare the two classical Oosterhoff populations with two GCs located in the Galactic bulge and we discuss the mean physical parameters of the Oosterhoff groups. Section \ref{sec:SpatDist} discusses the spatial distribution of stars in our sample. We summarize our results in Sect. \ref{sec:Conclus}.

\section{Sample selection} \label{sec:SamSel}
In order to study the Oosterhoff dichotomy in the Galactic bulge, we combined data from two large photometric surveys monitoring the area around the Galactic center, OGLE-IV and VVV. The OGLE survey monitors the Galactic bulge in optical wavelengths ($V$ and $I$-bands), while VVV conducted observations in the near-infrared ($Z$, $Y$, $H$, $J$, and $K_{s}$-bands). Therefore, the combination of both datasets allowed us to cover a broad wavelength range. 

To ensure homogeneity and detection completeness of our sample, we employed cuts to the sample of OGLE-IV variables and the resulting sample was then cross-matched with the data from the VVV survey. For the purpose of this study, we thus considered only variables with data in both surveys. In the Galactic bulge, OGLE-IV observed over 38\,000 RR~Lyrae stars of which more than 27\,000 are fundamental-mode pulsators \citep{Soszynski2014,Soszynski2017}. For the majority of objects the fourth OGLE release contains photometric data for individual stars, mean magnitudes in the $V$ and $I$-bands, pulsation periods, peak-to-peak amplitudes and Fourier coefficients ($R_{21}$, $\varphi_{21}$, $R_{31}$, $\varphi_{31}$). To examine the Oosterhoff dichotomy in the Galactic bulge we used only RRab stars for which all aforementioned Fourier coefficients and mean magnitudes were available in both filters. We also omitted variables that belong to GCs (\texttt{gc.dat}\footnotemark), have uncertain identification (\texttt{remarks.dat}\footnotemark[\value{footnote}]),\footnotetext{Files provided by the OGLE-IV survey from their ftp data access.} and stars exhibiting additional periods \citep{Smolec2016,PrudilSmolec2017}. In addition, we employed cuts based on the color-magnitude diagram (CMD) using the selection criteria suggested by \citet{Pietrukowicz2015}:

\begin{gather} 
I \leq 1.1 (V - I) + 16.0 \\
I \geq 1.1 (V - I) + 13.0 \\
V - I > 0.3 \\
I < 18.0. 
\end{gather}
These cuts were performed in order to remove stars with unreliable colors (bluer than the bulk of RR Lyrae pulsators in the Galactic bulge) and to improve the homogeneity of our selected sample \citep[see top panel of fig.~1 in][]{Pietrukowicz2015}. In the end, 11\,888 variables passed the aformentioned thresholds.

\subsection{Removing stars with the Blazhko effect} \label{sec:Classi-Blaz}

As shown by \cite{PrudilOO2018}, the Blazhko effect can hide the existence of the Oo\,I and II populations in the Galactic bulge. One way to detect the Blazhko effect is the period analysis of individual stars, which in our case would be enormously time-consuming due to our sample size. Therefore, we combined two different approaches to remove stars from our sample exhibiting the Blazhko effect. First, we removed the fundamental-mode pulsators that were previously identified to show the Blazhko effect \citep[over 40\,\% variables in the sample studied by][from hereon referred to as PS17]{Prudil2017}. We employed a 3$\sigma$ clipping based on the variables' position in the $\varphi_{31}$ vs. $R_{31}$ and $\varphi_{21}$ vs. $R_{31}$ planes using equations 5 and 6 from PS17. 

Secondly, we used supervised machine learning to classify stars with Blazhko modulation. We used the data of RR~Lyrae stars classified as modulated or non-modulated by PS17 (over 8\,000 variables) to train various classifiers implemented in \texttt{scikit-learn} \citep{Pedregosa2011}, a library of the Python programming language. As input features we used the pulsation periods, total amplitudes, and a combination of the Fourier coefficients ($R_{21}$, $R_{31}$, $\varphi_{21}$, $\varphi_{31}$).

The classifier's performance can be hampered by under- or overfitting the training sample, which can result in poor performance on the test dataset. We can optimize the performance via cross-validation, i.e., by dividing the already classified dataset into two parts: a training sample, and the validation sample. The model's parameters are fitted to the training data, while its hyper-parameters (i.e., parameters that govern the model complexity or the fitting procedure) are optimized on the validation data. The tuning of hyper-parameters was carried out using the module \texttt{GridSearchCV} with 10-fold cross-validation by maximizing one of the metrics (Average precision), from the aforementioned library. In this procedure, the classified data from PS17 were randomly shuffled and then divided 10 times into training and validation sets with 9:1 ratio in a way that every star is used in the a validation set only once. The resulting model was then applied to the test set that was not used in building and optimizing the model in order to evaluate the performance of the classifier. 

Here we provide a brief overview of the tested classifiers:
\begin{itemize}
\item The \texttt{Random forest} (RanForest) is a bootstrapping algorithm based on Decision tree models. Decision trees are accumulated into the final forest to determine the classification value. The Random forest builds a model with different samples and various initial variables. This process is repeated several times before the final model and prediction are made. Therefore, the final prediction is a function of the prediction of each model.
\vspace{0.2cm}
\item The \texttt{Support vector machines} (SVMk) project initial variables into a high-dimensional space. Then the algorithm searches for a linear hyperplane to separate the training groups with the largest margin. Support vectors, in this case, are the data points located nearest to this hyperplane. Their removal would alter the location of the hyperplane and result in a different classification of the training data.
\vspace{0.2cm}
\item The \texttt{Multi-layer perceptron} (MLP) is an artificial neural network mimicking the pattern of a biological neural network. It contains at least three layers, the input, hidden, and the output layer. The number of neurons (nodes) in each hidden layer and the number of neurons (nodes) in each of them are hyper-parameters of the model. Each individual neuron is connected to all other neurons of the surrounding layers.
\end{itemize}
We refer the interested reader to the book by \citet{Ivezic2014} for a general and more thorough overview of machine learning and the classifiers. 

\begin{figure}
\includegraphics[width=\columnwidth]{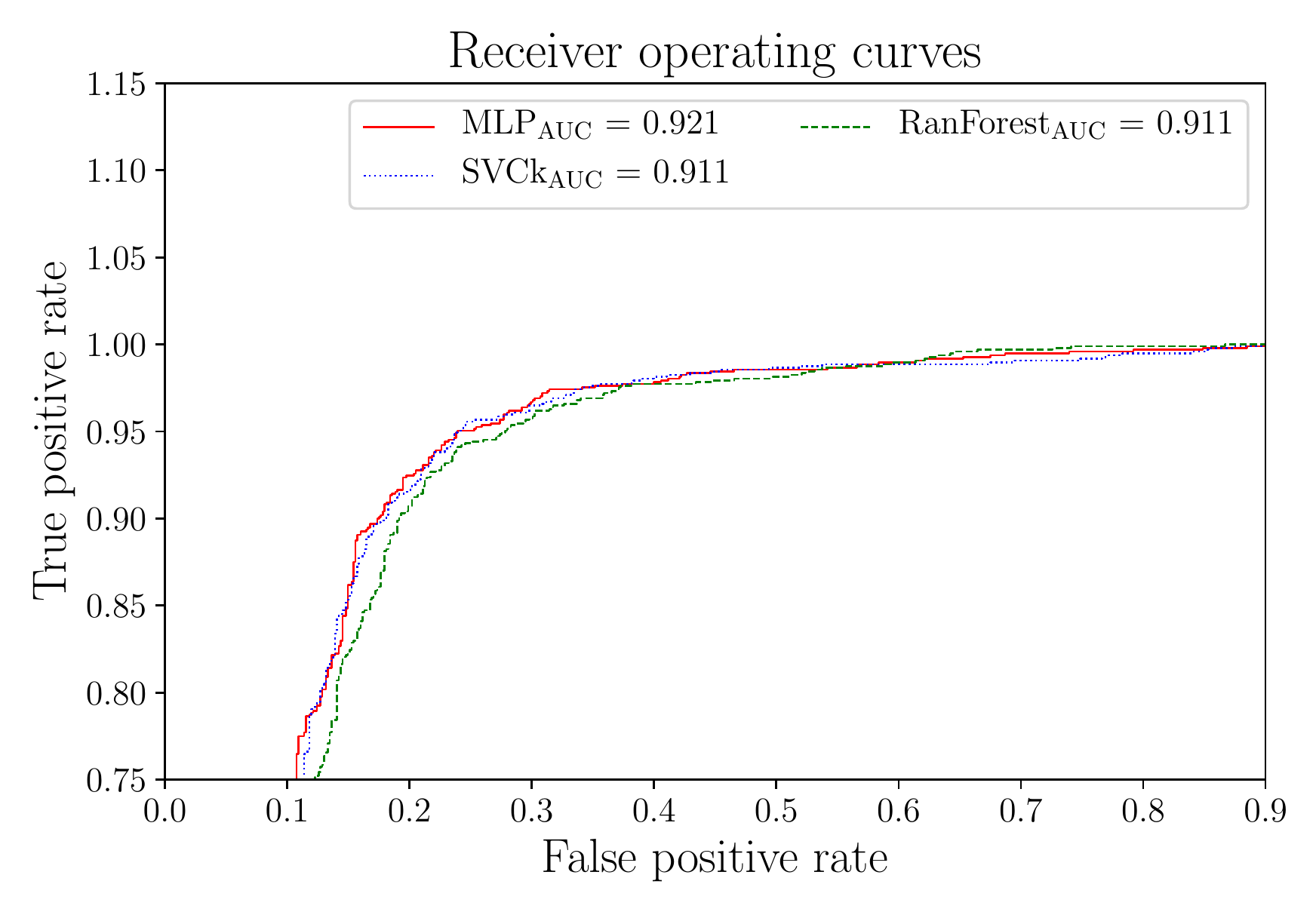}
\caption{Comparison of tested classifiers against the test sample. To increase clarity, we choose to show only that region in which the reader can easily distinguish differences between the tested classifiers. This figure shows receiver operating curves with lines (green, red and blue) representing the RanForest (dashed), MLP(solid) and SVMk (dotted). The MLP performs better in comparison with other classifiers, which is also reflected in the AUC listed in the legend for the individual classifiers.}
\label{fig:ClassMLcomparison}
\end{figure}

To select which classifier would serve best for our classification problem we used several performance metrics:

\begin{itemize}
\item Average precision 
\item Precision 
\item Recall
\item Accuracy score
\item Area under curve (under receiver operating curve)
\end{itemize}
where the average precision is the area under the precision-recall curve. Precision and recall are described by the following equations:

\begin{gather} 
\text{precision} = \frac{\text{TP}}{\text{TP + FP}},\\
\text{recall} = \frac{\text{TP}}{\text{TP + FN}},
\end{gather}
where the precision stands for the ratio between true-positives (TP) and the sum of TP and false-positives (FP). The precision (sometimes referred to as purity), represents how likely the classifier will be correct in giving a positive example. The recall (or completeness/sensitivity) represents the ratio between TP and the sum of TP and false-negatives (FN) and stands for a probability that a positive example will be retrieved by the classifier. To summarize, high precision gives a lower number of false examples and high recall gives high completeness of the retrieved positive examples. The accuracy score (ACC) is given by equation:

\begin{equation} \label{eq:AccScore}
\text{accuracy score} = \frac{\text{TP + TN}}{\text{P + N}},
\end{equation}
and represents the ratio between the sum of true positives and true negatives (TN) and the total population of examples (P - positive and N - negative). The area under the curve (AUC) stands for the area under receiver operating curves. The receiver operating curve (ROC) is a dependence between the true positive rate against the false positive rate (see bottom panel of Fig.~\ref{fig:ClassMLcomparison}). 

In Table~\ref{tab:ClasML} we list various performance metrics of the tested classifiers. The highest and second highest values for a given criterion are marked by bold-face and italics, respectively. The top classifier with the best performance was MLP (red line in Fig.~\ref{fig:ClassMLcomparison}), with the following hyper-parameters\footnote{For a thorough documentation of individual hyper-parameters see the \texttt{scikit-learn} description of the MLP classifier at \url{http://scikit-learn.org/stable/modules/generated/sklearn.neural_network.MLPClassifier.html}}:
\small
\begin{lstlisting}
parameters = {Hidden layers = (100,75,50,25),
               Activation function = 'tanh', 
               Learning rate = 'constant',
               Solver = 'lbfgs', 
               Regularization term $\alpha$: 0.75}
\end{lstlisting}  
\normalsize
Therefore, we used as a training sample stars in which the Blazhko effect was already identified by PS17 for the MLP classifier with optimized hyper-parameters to train the model that was applied to our selected sample. Stars identified by the classifier as non-Blazhko stars (8\,339 RR~Lyraes) remained in the sample, while stars classified as Blazhko candidates were removed. We note that the majority of the stars marked as modulated were short period variables associated with the Oo\,I population. 

\begin{table}
\centering
\caption{Performance of tested classifiers. Column 1 lists the classifiers: MLP, RanForest, SVMk. Column 2 contains average precision, columns 3 and 4 provide values for the precision and recall, respectively. Columns 5 and 6 contain the accuracy score (ACC) and area under the curve (AUC), respectively. Bold and italic fonts mark values with the best and second-best performance for each tested criterion.}
\begin{tabular}{llllll} 
\hline
Classifier                    & $\left\langle \rm Prec.  \right\rangle$    & Prec. & Recall  & ACC & AUC   \\ \hline 
RanForest                     & \textit{0.828} & \textit{0.842} & \textit{0.944} & \textit{0.862} & \textit{0.911} \\
MLP                           & \textbf{0.845} & \textbf{0.863} & 0.934 & \textbf{0.873} & \textbf{0.921} \\
SVMk                          & 0.816 & 0.825 & \textbf{0.961} & 0.856 & 0.911 \\
\hline 
\end{tabular}
\label{tab:ClasML}
\end{table}

\section{Distance determination and separation of the Oosterhoff groups} \label{sec:DistOogr}

To study the spatial distribution of the Oosterhoff groups in the Galactic bulge we first calculated their distances using period-metallicity-luminosity (P-M-L) relations, and then separated the Oosterhoff populations based on their position in the period-amplitude (\textit{P-A}) diagram.

\subsection{Metallicities and distances of RRab stars} \label{DisRRstars}

To determine the distance of RRab stars we took advantage of our broad range in color for individual RR~Lyrae stars in estimating their extinction, and then used $K_{s}$-band mean magnitudes from the VVV survey. In the near-infrared, the period-luminosity-metallicity relations for RR~Lyraes follow rather tight sequences, slightly less dependent on metallicity, and relations between absolute magnitudes and pulsation periods are known in various passbands \citep[see][]{Catelan2004,Alonso-Garcia2015,Marconi2015}.

For the optical data we first decomposed light curves using the Fourier series for the $I$-band photometric data: 

\begin{equation} \label{eq:FourierSeries}
m\left ( t \right ) = A_{0}^{I} + \sum_{k=1}^{n} A_{k}^{I} \cdot \text{cos} \left (2\pi k \vartheta + \varphi_{k}^{I} \right ),
\end{equation}
where $\varphi_{k}^{I}$ stands for phases, and $A_{k}^{I}$ represents amplitudes. The phase function $\vartheta$ is defined as follows: $\left(HJD-M_{0}\right)/P$, where the time of observation is denoted by $HJD$ (Heliocentric Julian Date), $M_{0}$ marks the time of maximum brightness and the pulsation period is represented by $P$. The $n$ represents the degree of the fit and was iteratively tailored for each light curve. First, we started the calculation with the third degree and increased the degree every time when the condition $A_{k}/\sigma_{k} > 4$ was met. Once this condition was no longer valid, we removed the outliers from the residuals, detrended the data using a low-degree polynomial, and started the iteration again. Stars for which we were unable to find a light curve solution including $\varphi_{3}^{I}$ was removed from the sample (198 stars). We note that the VVV survey provides only aperture photometry and not preferable point-spread function photometry. Thus, for each star, we selected an appropriate aperture based on the minimum value of a cost function for the fitted model on each light curve.

The obtained Fourier coefficients served to calculate the metallicity for the remaining 8\,141 stars using equation 3 from \cite{Smolec2005} (on the \cite{Jurcsik1995}, JK scale)\footnote{In the supplementary material for this paper we also provide transformed metallicities on the Carretta scale \citep{Carretta2009} using eq.~6 from \cite{Hajdu2018}.}:

\begin{equation} \label{eq:metallicity}
\text{[Fe/H]}_{\rm JK} = -6.125 - 4.795 P + 1.181 \varphi_{31} + 7.876 A_{2}.
\end{equation}
To properly account for errors in the metallicity we re-analysed relation 3 from \citet{Smolec2005} and calculated the covariance matrix for the three-parameter solution:
\begin{equation}
\text{Cov} = 
\begin{bmatrix}
0.6914 & -0.0632 & -0.0907 & -1.1951 \\
-0.0632 & 0.0813 & 0.0024 & 0.0428 \\
-0.0907 & 0.0024 & 0.0127 & 0.1430 \\
-1.1951 & 0.0428 & 0.1430 & 3.1552 
\end{bmatrix}
\end{equation}
This yields errors of on average 0.06\,dex. With a dispersion of $\sigma=0.14$\,dex from the \cite{Smolec2005} relation, the resulting average error in metallicity was 0.20\,dex. The calculated metallicities were rectified using a customized \textit{pyrime}\footnote{\url{https://github.com/gerhajdu/pyrime}} code from \cite{Hajdu2018} updated for our own Oosterhoff classification (for our definition see Subsection~\ref{SepOogroupsSubSec}). The final metallicities were then converted to $\text{log}Z=\text{[Fe/H]}_{\rm JK}-1.772$. We used the conversion factor $-1.772$ based on the solar metallicity determined by \cite{Grevesse1998} and a helium mass fraction of $Y=0.2485$. The log$Z$ and [Fe/H] were subsequently used to determine absolute magnitudes in the $I$ and $K_{s}$-bands using relations from \citet{Catelan2004} and \citet{Muraveva2018}, respectively, transformed into the VVV photometric system\footnote{For calculation the $M_{K_{s}}$ magnitudes we transformed our metallicities to the \citet{Zinn1984} metallicity scale using the relation from \citet{Papadakis2000}.}:

\begin{gather} \label{eq:Mi-Mk}
M_{I} = 0.471 - 1.132 \cdot \text{log}P + 0.205 \cdot \text{log}Z \\
M_{K_{s}} = -0.8481 - 2.5836 \cdot \text{log}P + 0.17 \cdot \text{[Fe/H]}_{\rm JK}.
\end{gather}
We note that we tested several theoretical relations for the absolute magnitude in $K_{s}$ band. We found, similarly to \citet[][their fig. 15]{Marconi2015} or \citet[][their table 3]{Muraveva2015}, non-negligible differences between different versions of the same relation, suggesting the presence of systematic uncertainties at the $\sim 0.1$~mag level. In the end, we decided to use the relation from \citet{Muraveva2018}, which is based on the latest (i.e. Data Release 2) parallaxes from the Gaia space telescope and is model-independent.

Absolute and apparent magnitudes then served to estimate the reddening vector. First, we omitted stars with $E\left ( I - K_{s} \right ) < 0.5$ and $E\left ( I - K_{s} \right ) > 2.5$ to limit the possible incompleteness at the bright and faint ends of our sample. We then applied an average boxcar filter on the $I-M_I$ vs. $E\left ( I - K_{s} \right )$ distribution with varying box width (from 300 to 800 points) and a step size (from 10 to 299 points) and linearly fitted each binned distribution. We took into account errors in $I - M_{I}$. As weights for the fit, we used the standard deviation of the $I - M_{I}$ distribution in the individual bins. The final coefficients of the linear fit were selected based on the median value of the distribution for the first coefficient. The results of this procedure are shown in Fig.~\ref{fig:IstvanFit}. For a derivation of the extinction in the $K_{s}$-band, we then followed:

\begin{figure}
\includegraphics[width=\columnwidth]{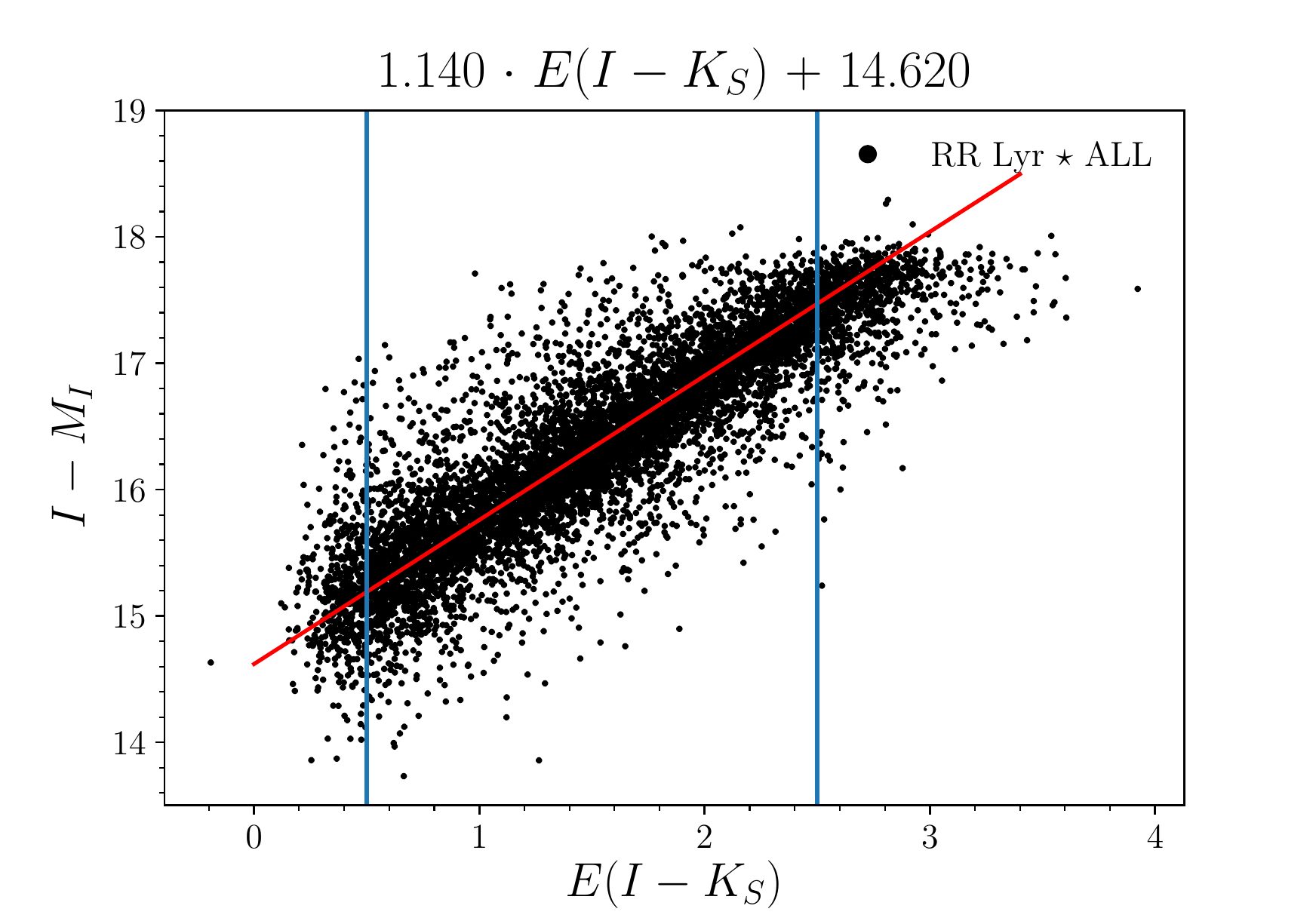}
\caption{The linear dependence between reddening $E\left ( I - K_{s} \right )$ and difference of apparent and absolute magnitude. The black dots represent stars from our sample together with a red line representing the reddening vector and blue lines outlining the applied boundaries.}
\label{fig:IstvanFit}
\end{figure}

\begin{gather} \label{eq:KI-EKi-Ak}
E\left ( I - K_{s} \right ) = \left ( I - K_{s} \right ) - \left ( I - K_{s} \right )_{\rm 0} \\
A_{K_{s}} = 0.14 \cdot E\left ( I - K_{s} \right ).
\end{gather}
Lastly, the distances $d$ (in pc) to the individual RRab stars were detemined from the distance modulus:

\begin{equation}
d = 10^{1 + 0.2 \cdot \left ( K_{s} - M_{K_{s}} - A_{K_{s}}\right )}.
\end{equation}
The errors in the distances were estimated as well. We assumed 0.02\,mag as the average errors for the $I$ and $K_{s}$-band light curves \citep{Dekany2013,Udalski2015}. With the individual error estimates for the metallicities, the errors in distance vary around a mean of 0.19\,kpc.

\subsection{Separation of Oosterhoff groups} \label{SepOogroupsSubSec}

Traditionally, the separation of the two Oosterhoff groups can be achieved in the \textit{P-A} plane using linear/polynomial relations (see \citet{Clement1999} and \citet{Cacciari2005}, respectively). These relations are based on the fundamental-mode RR~Lyrae stars in GCs and the overall position of the Oosterhoff type I locus. As shown by \citet{Kunder2009}, the locus of Oosterhoff type I variables in the Galactic bulge is shifted to lower pulsation periods approximately by 0.02\,days, in comparison with the archetype of Oo\,I clusters, M3. This effect is due to the difference in metallicity of the aforementioned systems \citep[-1.5\,dex for M\,3 and -1.0\,dex for the Galactic bulge, respectively,][]{Cacciari2005,Pietrukowicz2015}. Due to this effect, we tailored our own polynomial relation based on the selected sample in the $V$, $I$, and $K_{s}$-bands.   

The OGLE-IV survey provides $V$-band data for our sample stars, but the data in this passband are rather sparse. Specifically, more than two-thirds of the sample stars have less than 86 observations in the $V$-band. Therefore, we defined a relation for $V$-band amplitudes of RR~Lyraes using the $I$-band photometry. We selected stars from our sample that have at least 86 observations in the $V$-band (in total 2\,671 variables). Using a Fourier decomposition (see Eq.~\ref{eq:FourierSeries}) of fourth degree we calculated amplitudes for the selected RR~Lyraes in the passband $V$. The $V$-band amplitudes were compared with their $I$-band counterparts and a transformation equation was obtained in the following form:

\begin{equation}
\text{Amp}^{V} = 1.6130 \pm 0.0060 \cdot \text{Amp}^{I},
\end{equation}
where Amp$^{V}$ and Amp$^{I}$ stand for the amplitudes in the aformentioned passbands. We set the intercept to zero to preserve the physical representation of the calculated $V$-band amplitudes. We note that this relation between amplitudes can depend on metallicity since amplitudes appeared in some of the empirical relations for metallicity \citep[e.g. see eq. 6 in][]{Sandage2004}. We tested this in several metallicity bins and found only a negligible discrepancy between individual bins.

\begin{figure*}
\includegraphics[width=\columnwidth]{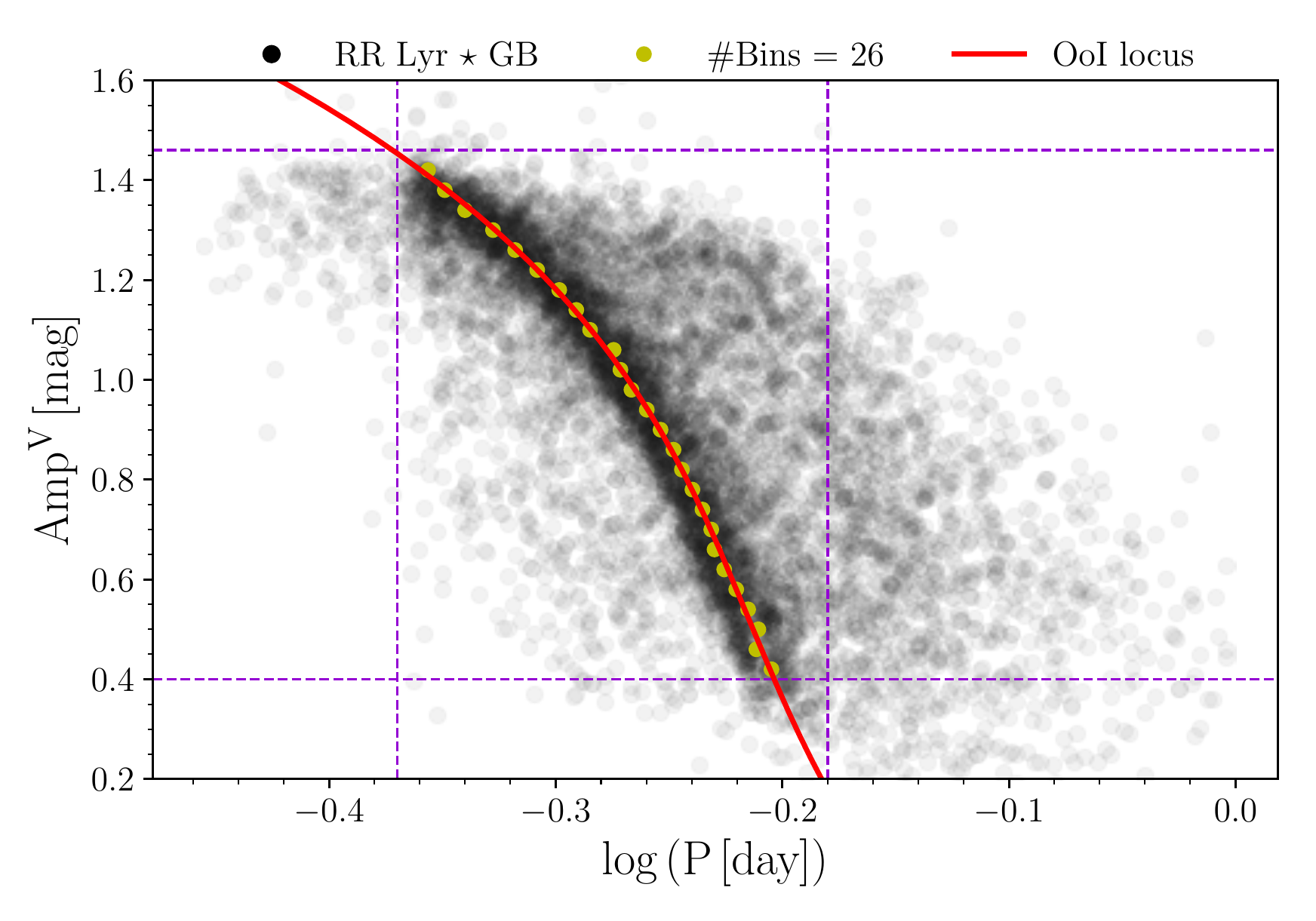}  
\includegraphics[width=\columnwidth]{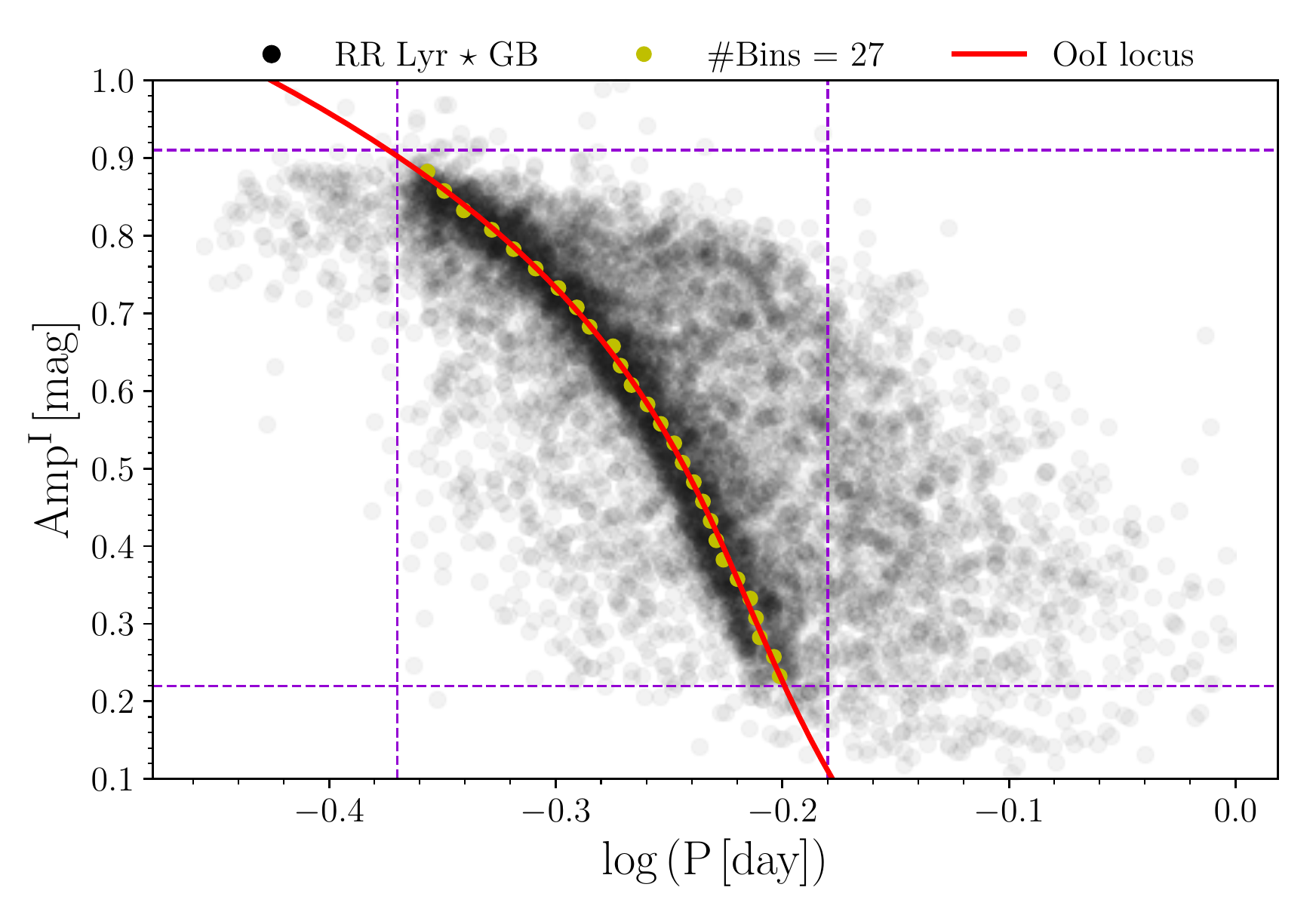} \\
\includegraphics[width=\columnwidth]{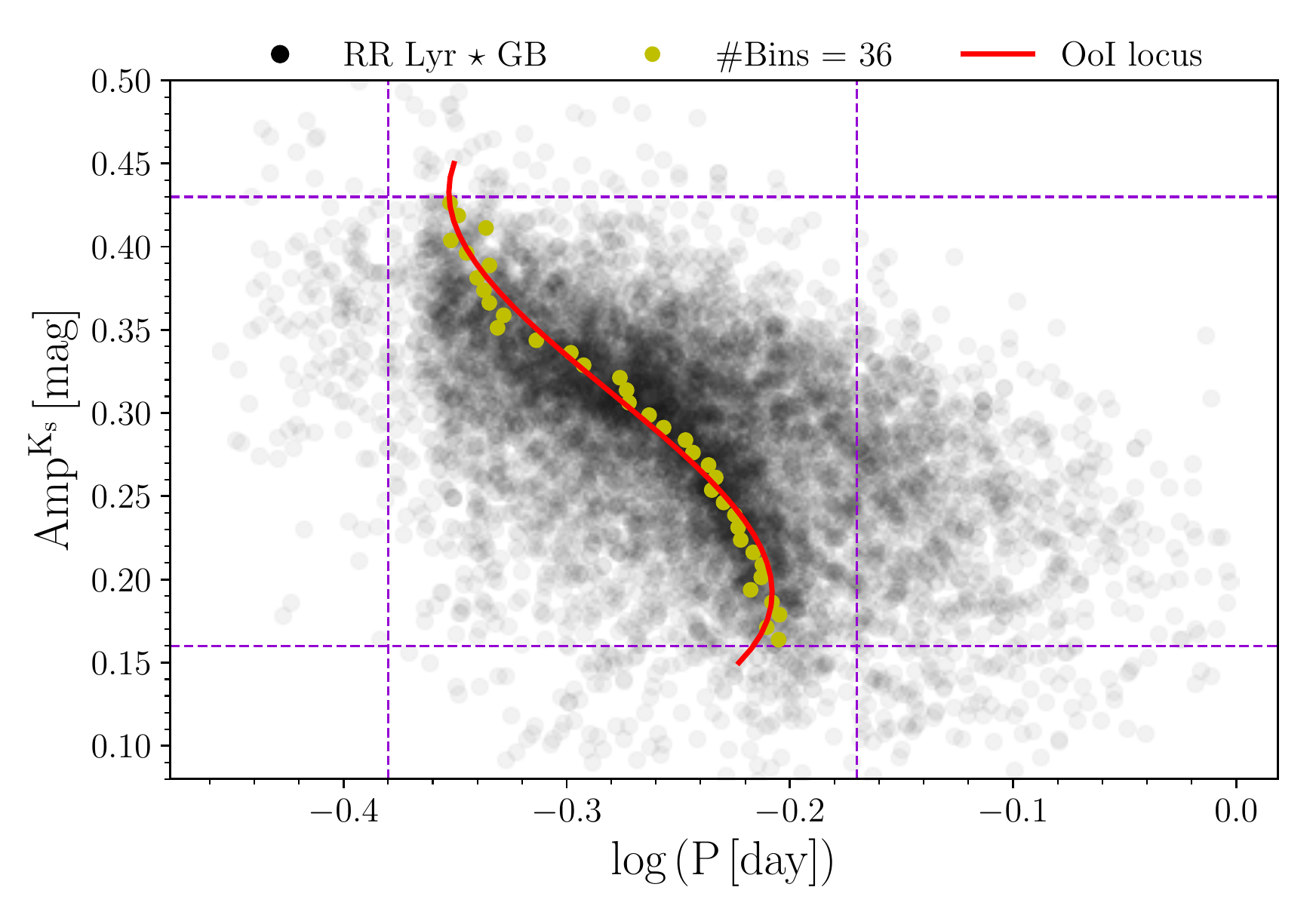}
\includegraphics[width=\columnwidth]{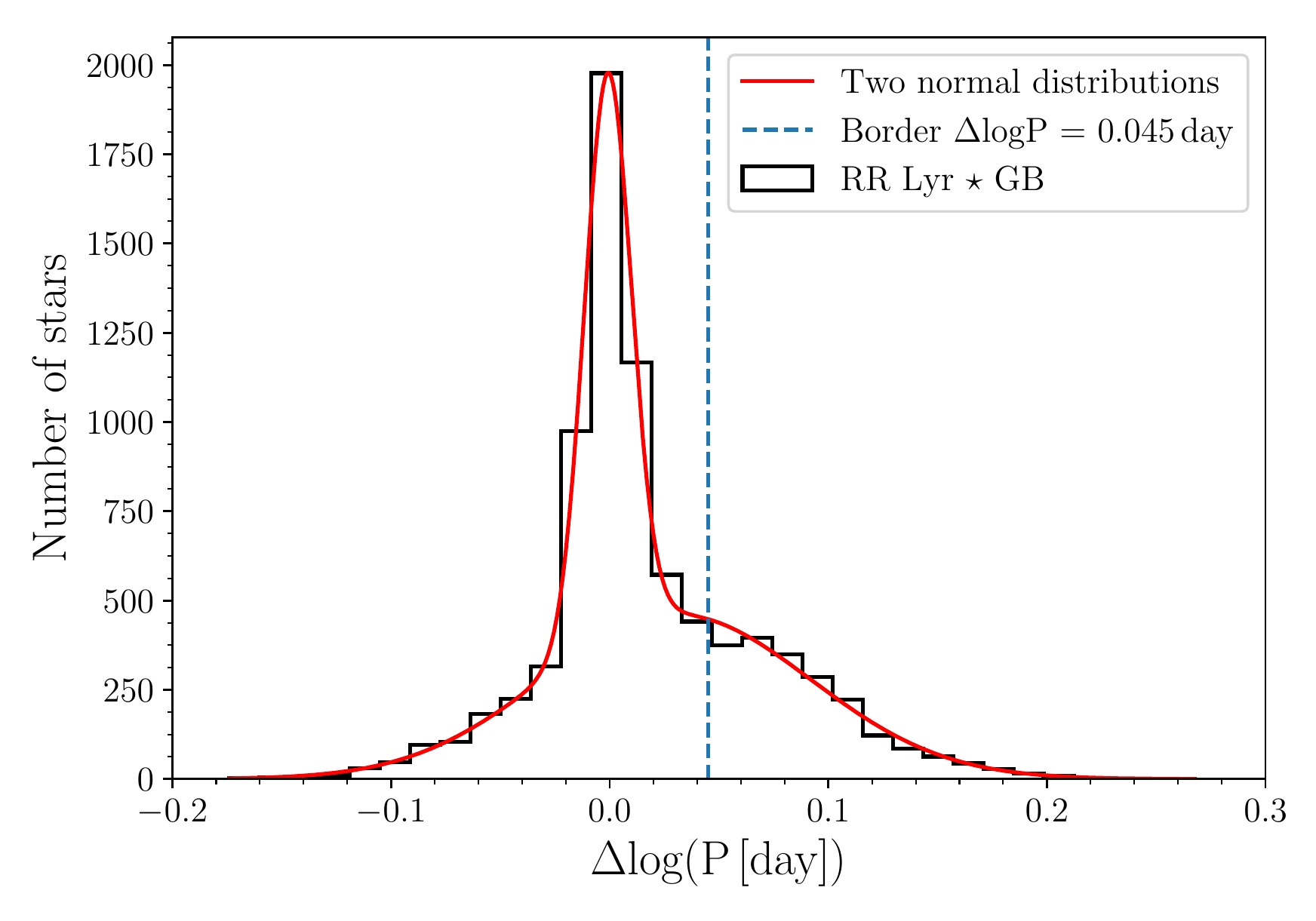}

\caption{The \textit{P-A} diagrams for selected filters (top left panel: $V$, top right panel: $I$, bottom left panel: $K_{s}$) of studied RR~Lyrae stars in the Galactic bulge (GB), and the distribution of period differences based on the defined relation for $I$-band amplitudes (bottom right panel). In the \textit{P-A} diagrams, the yellow circles stand for the binned Oo\,I locus. The total number of bins is listed in the legend of each plot. The violet dashed lines display a cut in order to select solely the Oo I progression in the \textit{P-A} diagrams for each passband individually. The black dots represent fundamental-mode RR~Lyraes, and the red line shows our third-degree polynomial equation for the Oo I locus. Lower right panel: The distribution of $\Delta$logP values, obtained using the third degree polynomial relation for the Oo\,I locus in the $I$-band amplitudes, is shown as a black line. The distribution was fitted with a sum of two Gaussian distributions (red line), the blue dashed line denotes the borderline between both Oosterhoff groups.}
\label{fig:Period-Amp-OO}
\end{figure*}

To define a polynomial relation that would adequately describe the locus of Oo\,I stars we proceeded in the following way. We binned our stellar sample based on their amplitudes, computed kernel density estimates (KDE) as a function of the period for each amplitude bin and computed the maximum in each individual bin. These maxima were then fitted with third and second-degree polynomial relations. We proceded in the same way in all available colors to define the Oo\,I progression. The number of amplitude bins were manually optimized for each passband. The results of this procedure are shown in Fig.~\ref{fig:Period-Amp-OO} and the individual coefficients are listed in Table~\ref{tab:coofDifColors} for all three passbands.

\begin{table}
\centering
\caption{Tables of coefficients for third and second-degree polynomials describing the Oo\,I locus. Each table contains a list of filters (column 1). The top table contains coefficients for the third-degree polynomial (columns 2 - 5) and column 6 contains the $\sigma$ of the fit. The bottom table shows the coefficients for the second-degree polynomial (columns 2 - 4) and the $\sigma$ of the fit (column 5).}
\label{tab:coofDifColors}
\begin{tabular}{lccccc}\hline 
\multirow{2}{*}{Filter} & \multicolumn{5}{c}{log$P$ = $a_{1}$ $\cdot$ Amp$^{3}_{i}$ + $a_{2}$ $\cdot$ Amp$^{2}_{i}$ + $a_{3}$ $\cdot$ Amp$_{i}$ + $a_{4}$} \\
                  & $a_{1}$                  & $a_{2}$                  & $a_{3}$                & $a_{4}$                 & $\sigma$       \\ \hline
$V$                 & -0.084 & 0.129 & -0.157 & -0.156 & 0.0016 \\
$I$                & -0.324 & 0.289 & -0.229 & -0.160 & 0.0015 \\
$K_{s}$                 & 21.356 & -20.004 & 5.340 & -0.647 & 0.0066  \\ \hline \hline
\multirow{2}{*}{Filter} & \multicolumn{5}{c}{log$P$ = $a_{1}$ $\cdot$ Amp$^{2}_{i}$ + $a_{2}$ $\cdot$ Amp$_{i}$ + $a_{3}$} \\
                  & $a_{1}$                  & $a_{2}$                  & $a_{3}$                & $\sigma$ &        \\ \hline
$V$                 & -0.102 & 0.042 & -0.209 & 0.0025 & \\
$I$                & -0.253 & 0.051 & -0.204 & 0.0025 & \\ 
$K_{s}$                 & -1.104 & -0.003 & -0.168 & 0.0106 &   \\ \hline
\end{tabular}
\end{table}

\begin{figure} 
\includegraphics[width=\columnwidth]{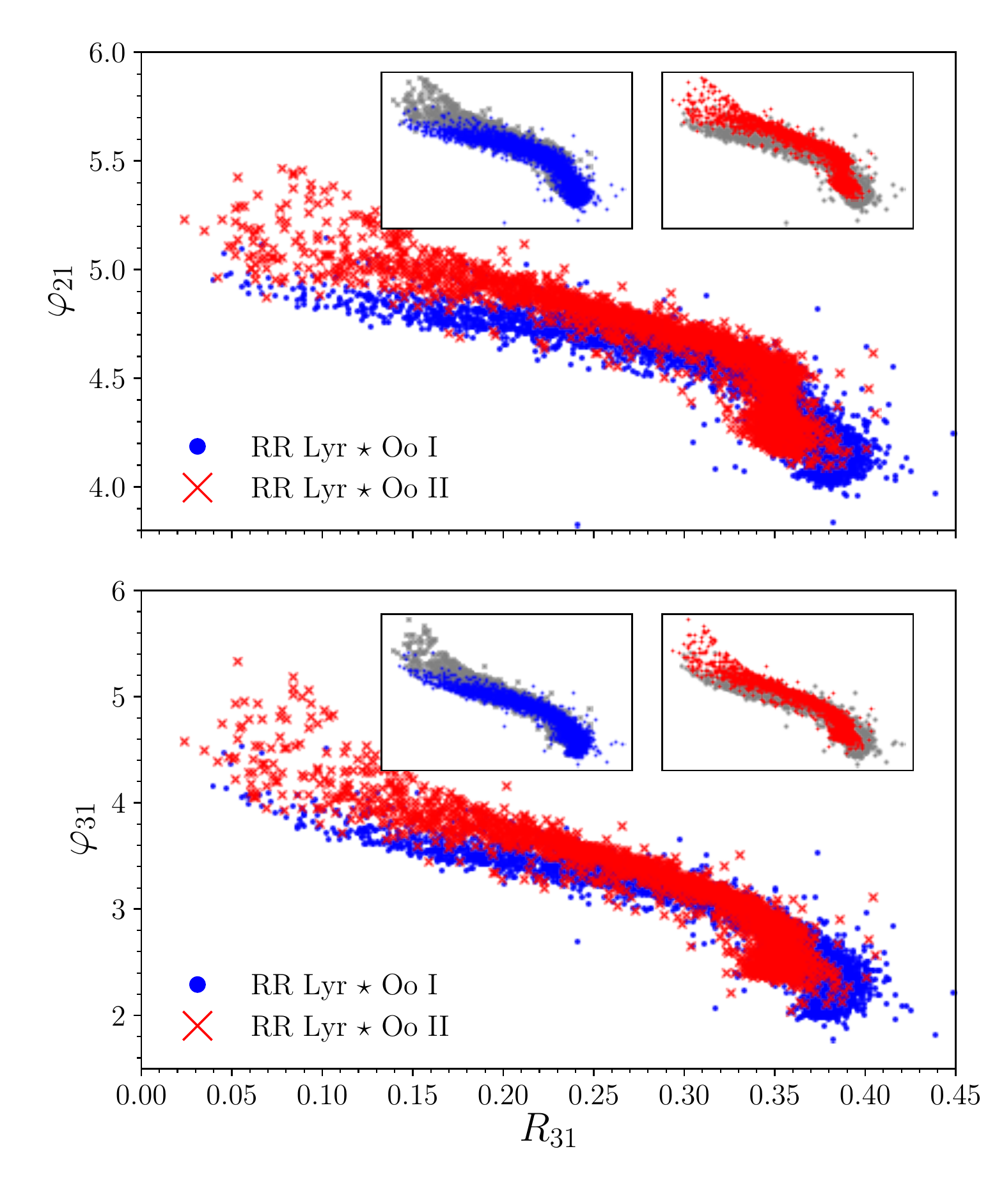}
\caption{Distributions of our non-Blazhko bulge RRab sample on the $\varphi_{21}$ {\em vs} $R_{31}$ (top) and the $\varphi_{31}$ {\em vs} $R_{31}$ (bottom) planes. In both panels, blue dots represent Oo\,I stars, while red crosses stand for Oo\,II variables. The insets in both panels display the separated Oosterhoff groups highlighted with blue/red colors and the rest of the sample marked by gray points to enhance the visibility of the two-hook structure.}
\label{fig:R31-phi21phi31}
\end{figure}

In the \textit{P-A} diagrams, we can distinguish the Oo\,I locus, but an Oo\,II locus (expected toward the right edge of the distributions) is not visually apparent. In the bottom right panel of Fig.~\ref{fig:Period-Amp-OO} we show the distribution of the difference between the observed and calculated pulsation periods ($\Delta$logP, based on the period-amplitude relations from Tab.~\ref{tab:coofDifColors}). We see a sign of a bimodal distribution with a peak at zero (the Oo\,I locus) and an enhanced tail on the positive side (consistent with Oo\,II stars). Thus, we decided to apply the conditions prescribed by \citet{Miceli2008} to separate between the two Oosterhoff groups while using a fixed threshold with respect to the third-degree polynomial fitted to the $I$-band amplitudes:

\begin{equation} \label{eq:MiceliSplit}
\Delta P = P_{\rm ab} - P_{\rm Oo I} \begin{Bmatrix}
\mathrm{RRab}_{\rm Oo I} & \Delta P \leq 0.045\,\mathrm{days}\\ 
\mathrm{RRab}_{\rm Oo II} & \Delta P > 0.045\,\mathrm{days}
\end{Bmatrix}.
\end{equation} 
We note that the value of 0.045\,days is based on RR~Lyrae stars in GCs. We tested different values for the boundaries between both Oosterhoff populations ranging in $\Delta$log$P$ from 0.035 to 0.055. Values between 0.040 and 0.050 yield analogous results when compared with the separation of the two-hook structure in the top panel of Fig.~\ref{fig:R31-phi21phi31}. The uncertainty in this value is mostly due to the metallicity dependence of the Oo\,I locus.

The resulting separation based on $I$-band amplitudes yields 6\,086 variables associated with the Oo\,I group, and 2\,055 stars with the Oo\,II group. A similar separation using the $V$ passband yields almost the same classifications as the $I$-band data (the disagreement is around 1\,\%). This negligible difference was most likely caused by the transformation from $I$-band amplitudes into the $V$-band. On the other hand, the separation based on the infrared amplitudes yields different Oosterhoff classifications for 7.7\,\% of the sample. This discrepancy was probably caused by the large scatter in the \textit{P-A} diagram and the lack of a pronounced Oo\,I locus in the $K_{s}$-band. In the following, we will work with the Oosterhoff groups separated based on the $I$-band \textit{P-A} relation. 

Our sample shows two-hook structures in the $\varphi_{21}$ vs. $R_{31}$ and $\varphi_{31}$ vs. $R_{31}$ planes (see Fig.~\ref{fig:R31-phi21phi31}). These structures were identified as two Oosterhoff populations in the Galactic bulge by \citet{PrudilOO2018}. On the other hand, in the period-amplitude diagram we do not observe the locus of the Oosterhoff type II variables, but only a pronounced Oosterhoff type I locus (top-left panel of Fig.~\ref{fig:Period-Amp-OO}). 

In total, approximately 25\,\% of the stars in our sample belong to the Oo\,II group, which is almost identical to the 1:4 ratio found by \citet[][24\,\%, in a sample of 12\,227 fundamental mode RR Lyraes]{Drake2013} and \citet[][25\,\%, in a sample of 4\,067 RRab type stars]{Sesar2013} for the Galactic halo. However, the \textit{P-A} diagrams for the Galactic bulge and Galactic halo differ. In the Galactic halo, the Oo\,I locus fits the polynomial relation by \citet{Cacciari2005} due to its lower metallicity. In addition, the \textit{P-A} distribution of Galactic halo RR~Lyraes shows a more pronounced Oo\,II locus than in the bulge.

\subsection{The separation of the Oo groups by Manifold learning}

The separation of the Oosterhoff groups as described in Sec.~\ref{SepOogroupsSubSec} is based on external information (i.e., locus positions in GCs), which might be debatable due to the fact that the Oo\,II locus is not apparent on the \textit{P-A} plane for the bulge.. One can try and infer the Oo separation from the two-hook structure in $\varphi_{21}$/$\varphi_{31}$ vs. $R_{31}$ using clustering algorithms, but due to their tight entanglement, this could be challenging. 

Thus we decided to employ \texttt{Manifold-learning} from the \texttt{scikit-learn} library. The \texttt{Manifold-learning} is a type of unsupervised machine learning algorithms that perform non-linear dimensionality reduction and thus project data into a low-dimensional Eucledian space. We looked for the separation of the two Oosterhoff groups based on the same parameters that were used for the removal of modulated stars (i.e., pulsation periods, amplitudes, $R_{21}$, $R_{31}$, $\varphi_{21}$, $\varphi_{31}$ see Sec.~\ref{sec:Classi-Blaz}) and $\Delta$log$P$ calculated from the third-degree polynomial relation for $I$-band amplitudes. We used the \texttt{scikit-learn} implementation of the \texttt{Isometric Mapping (Isomap)} algorithm \citep{Tenenbaum2000}, which searches for a lower-dimensional setup while preserving geodesic distances between data points. 

\begin{figure*}
\includegraphics[width=504pt]{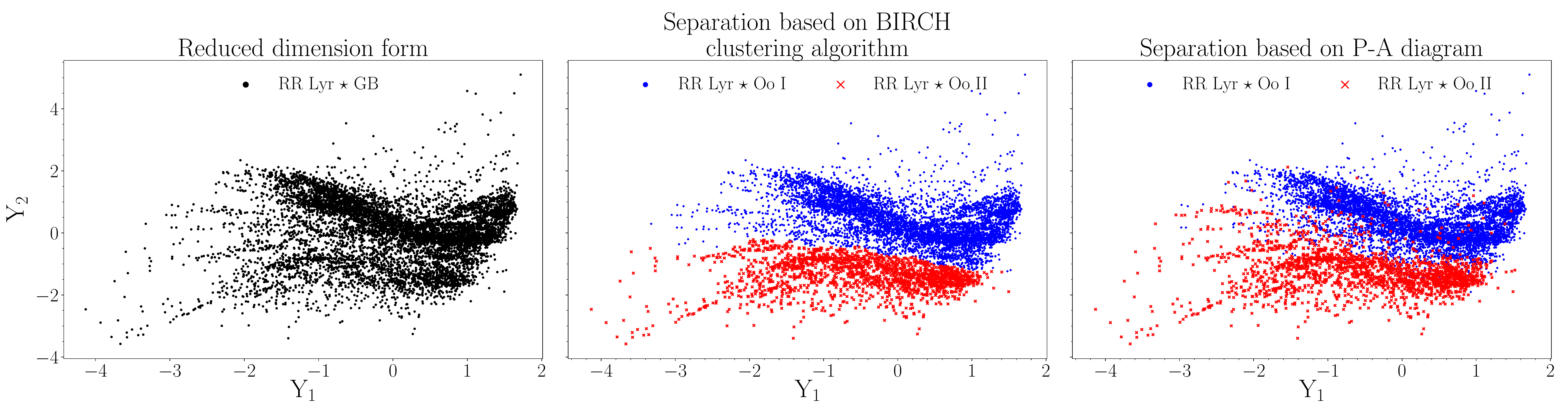}
\caption{Reduced dimensions (Y$_{1}$ and Y$_{2}$) based on the \texttt{Manifold Isometric Mapping} algorithm. The left-hand panel displays sample stars (black points) after dimension reduction with two structures representing the Oosterhoff populations in our data. The middle panel shows separated groups based on the \texttt{Birch} clustering algorithm with blue dots representing Oo\,I variables and red crosses standing for Oo\,II pulsators. The same color-coding applies also for the right-hand panel where both Oosterhoff groups are depicted based on the separation in the \textit{P-A} space.}
\label{fig:Manifold}
\end{figure*}

The results of the \texttt{Isomap} algorithm can be seen in the left panel of Fig.~\ref{fig:Manifold}, where we show the reduced dimensions of our RRab sample. We clearly see two rather well-separated groups in the left-hand panel of Fig.~\ref{fig:Manifold}. To separate Oosterhoff groups, we used the implementation of the \texttt{Birch} clustering algorithm from the aforementioned library. We used the entire stellar sample associated with the two Oosterhoff groups based on the \textit{P-A} diagram, and trained the clustering algorithm for the best performance, which separated both clusters (middle panel of Fig.~\ref{fig:Manifold}) and divided them consistently with the aformentioned Oosterhoff groups (see Sec.~\ref{SepOogroupsSubSec}). When compared to the original classification based on the period-amplitude diagram we found agreement in 95\,\% of the cases. The rest of the stars lie on the boundary between the Oo\,I and Oo\,II groups as divided on the basis of the period-amplitude diagram. 

In the left-hand panel of Fig.~\ref{fig:Manifold} we see a small clump between the identified Oo\,I and II groups. Closer inspection of the period-amplitude diagram shows that stars associated with this structure fall on the low-amplitude, long-period end of the Oo\,I group. In this subsection we showed that it is possible to identify the Oosterhoff dichotomy and separate the two Oosterhoff groups purely on the basis of the properties of the bulge RR~Lyraes and the location of the Oo\,I locus. This approach can be used in any system that does not show both Oo loci in the \textit{P-A} plane.

\section{Comparison with bulge GCs and physical parameters}\label{NGC6441-PhysiParam}

The OGLE-IV photometry contains information about 300 RR~Lyraes that are probable members of 15 GCs \citep[see the bottom-left panel of fig.~2 in][for their spatial distribution]{Soszynski2014}. We decided to compare our two Oosterhoff populations with two Galactic bulge GCs with the highest number of fundamental mode RR~Lyraes in OGLE-IV data\footnote{\url{http://www.astrouw.edu.pl/ogle/ogle4/OCVS/blg/rrlyr/gc.dat}}, NGC~6401 and NGC~6441. We excluded the globular cluster M\,54 from this comparison due to its location in the Sagittarius dwarf spheroidal galaxy. 

\begin{figure}
\includegraphics[width=\columnwidth]{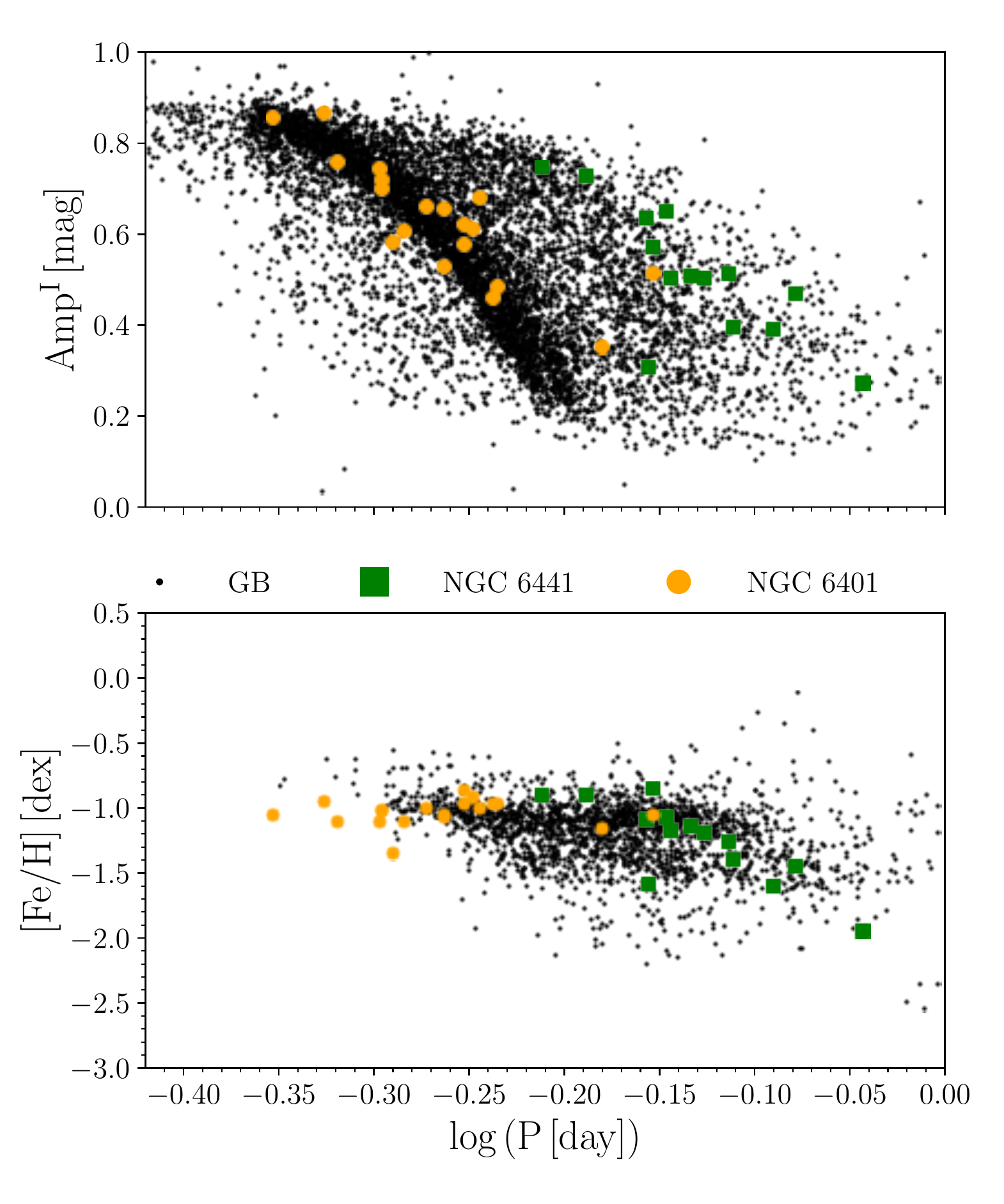} \\
\includegraphics[width=\columnwidth]{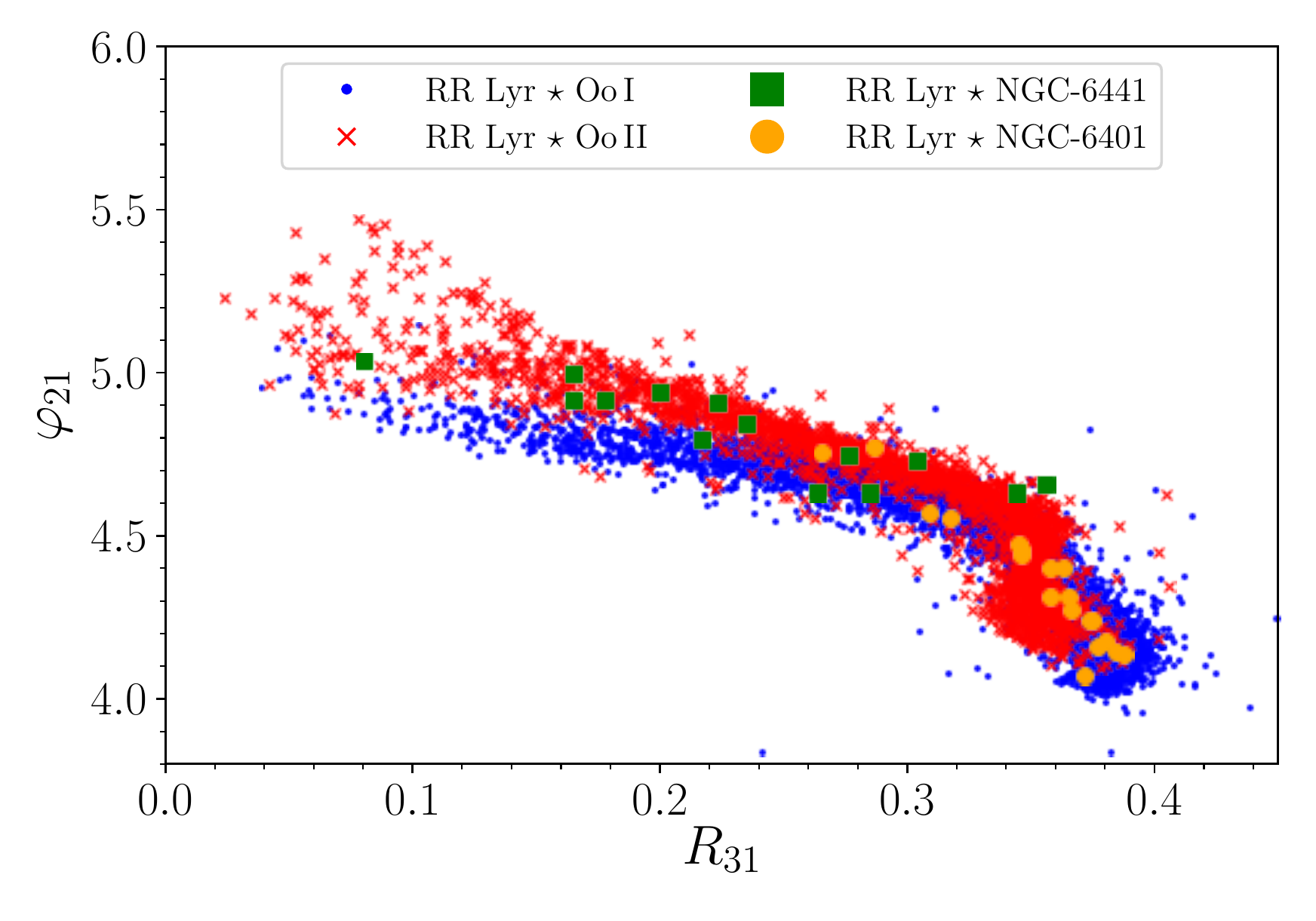}
\caption{The \textit{P-A} diagram (top panel), metallicity vs. period distribution (middle panel), and $\varphi_{21}$ vs. $R_{31}$ dependency (bottom panel). In all three figures, the orange circles and green squares represent RRab stars in the GCs NGC~6401 and NGC~6441, respectively. In the top panel the black dots represent all RRab variables from our selected sample while in the middle plot they stand for Oo\,II stars only. The bottom panel shows Oo\,I and Oo\,II variables (blue dots and red crosses, respectively).}
\label{fig:NGC-P-A}
\end{figure}

The globular cluster NGC~6401 is considered as a metal-rich globular cluster \citep[$\text{[Fe/H]}_{\rm JK}=-0.98$\,dex,][]{Tsapras2017}. We performed Fourier analysis using the $I$-band light curves and estimated the metallicity of the associated RRab variables using Eq.~\ref{eq:metallicity}, and removed potential modulated stars using the trained classifier (see Sec.~\ref{sec:Classi-Blaz}). This yields a median metallicity of $\text{[Fe/H]}_{\rm JK}=-1.02 \pm 0.05$\,dex, which is in good agreement with the average metallicity of RR~Lyraes in the Galactic bulge $\text{[Fe/H]}_{\rm JK}=-1.02\pm0.25$\,dex, \citep{Pietrukowicz2015}. Based on the positions of its RRab variables in the \textit{P-A} diagram, the $\text{[Fe/H]}$ vs. log$P$ dependence, and the $\varphi_{21}$ vs. $R_{31}$ distribution (orange circles in Fig.~\ref{fig:NGC-P-A}), the RR~Lyraes in NGC~6401 clearly belong to the Oo\,I group with a metallicity of the old bulge population. 

The globular cluster NGC~6441 is a larger system in comparison with NGC~6401. It is a known oddball in the Oosterhoff population scheme. Together with NGC~6388, these two clusters create their own group \citep[sometimes referred to as Oosterhoff type III,][]{Pritzl2000}. The fundamental-mode pulsators in NGC 6441 cluster occupy the long-period end of the \textit{P-A} diagram (green squares in Fig.~\ref{fig:NGC-P-A}), similarly to Oo\,II stars, yet they have a high mean metallicity of $\text{[Fe/H]}_{\rm JK}=-0.43$\,dex\footnotetext{Transformed from the \citet{Zinn1984} scale to [Fe/H]$_{\rm JK}$ using the relation from \citet{Papadakis2000}.} \citep{Clementini2005}. We performed a light curve decomposition on RRab variables associated with NGC~6441, and removed stars marked by the trained classifier as possible modulated variables. We detrended the data for individual pulsators (nearly all stars show trends in the mean brightness) and calculated their metallicities using Eq.~\ref{eq:metallicity}. Their median metallicity is $\text{[Fe/H]}_{\rm JK}=-1.19 \pm 0.23$\,dex, which does not agree with the spectroscopic measurements by \citet{Clementini2005}. We comment on this discrepancy later in this Section. The RRab stars associated with this cluster occupy similar regions in the \textit{P-A} diagram as some of the stars identified as the Oo\,II variables. This hints at the possibility that some Oo\,III stars may be hidden in our bulge RRab sample. Furthermore, \citet{Kunder2018} found eight RRab stars in up to three times the distance from the tidal radius of the NGC~6441, which share similar radial velocities as RR~Lyrae members of NGC~6441. This points towards the possible occurrence of stripped Oo\,III variables in the Galactic bulge among the Oo\,II variables.

In addition, the low-amplitude double-mode variables found by \citet{Smolec2016} with a dominant fundamental mode and long periods fall roughly in the region of Oo\,III variables. Most of these double-mode pulsators can be described using linear pulsation models with very high metallicity (around -0.5\,dex). Thus, we performed a thorough analysis of the frequency spectra of fundamental-mode variables in NGC~6441. We found that the object OGLE-BLG-RRLYR-03918 exhibits an additional peak in the frequency spectrum at $f_{x}=1.99974$\,c/d with S/N=4.16. The period ratio of this additional peak and the dominant mode is 0.720, therefore it lies in a region of the Petersen diagram \citep[see fig.~1 in][]{Smolec2016} where the aforementioned double-mode pulsators occur. We did not detect any signal at the combination frequency of the dominant mode and the additional signal ($f_x$). The $f_x$ is close to the integer value in the frequency spectra caused by Earth's rotation. Therefore, this star should be considered as merely a candidate for possible double-mode pulsators found by \citet{Smolec2016}.    

In the middle panel of Fig.~\ref{fig:NGC-P-A}, showing the $\text{[Fe/H]}$ vs. log$P$ dependence, we see that RRab variables from NGC~6441 mostly occupy the same region as the Oo\,II stars in our sample. But for the RR~Lyraes in a globular cluster, we would expect a very narrow distribution in metallicities. What we see here suggests that the RR~Lyrae metallicity relation (Eq.~\ref{eq:metallicity}) does not apply to Oo\,III RR~Lyraes. We see a large dispersion in metallicity for RR~Lyraes in NGC~6441 when compared with RRab variables in NGC~6401 (yellow circles). The bottom panel of Fig.~\ref{fig:NGC-P-A} further corroborates our conclusions that the fundamental-mode pulsators from NGC~6441 coincide with the tail of the Oo\,II stars from our sample, and therefore some Oo\,III variables may be hidden in our Oo\,II sample.

\subsection{Physical parameters}

Based on the photometric data we can also estimate some of the physical parameters of the selected variables. In this analysis we assume that the calibrations are equally valid for the Oo\,I and Oo\,II groups. To calculate effective temperatures we used equation 6 from \citet{Jurcsik2018}. From the Fourier analysis performed in \S~\ref{DisRRstars} we also estimated the rise time (time between the minimum and maximum light, from hereon RT) for individual variables. The calculated median physical parameters for both Oosterhoff groups can be found in Table~\ref{tsb:physparamOo}.  

\begin{figure} 
\includegraphics[width=\columnwidth]{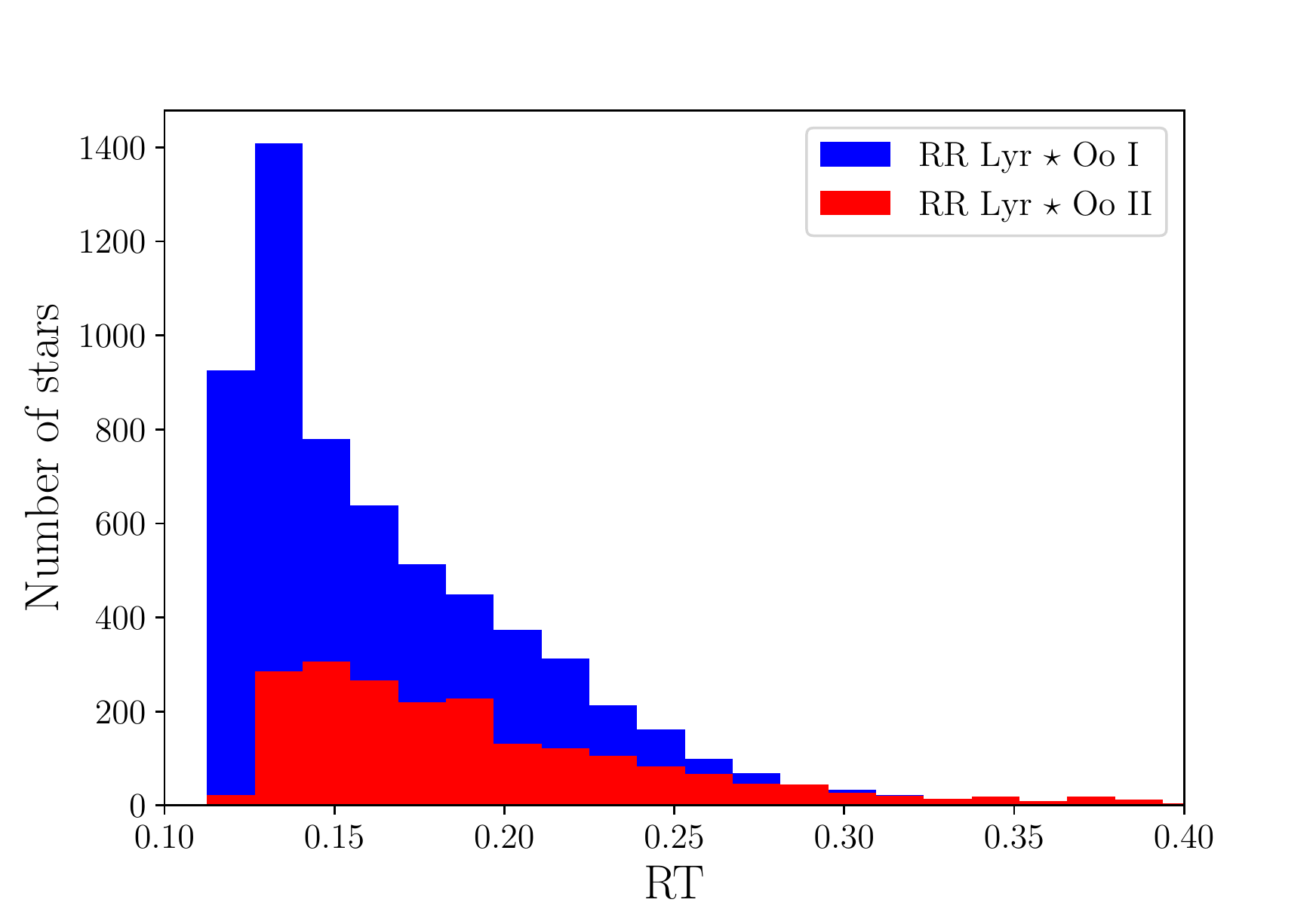}
\caption{The distribution of rise times (RT) for Oosterhoff groups from our sample. The Oo\,I and Oo\,II groups are represented by blue and red columns, respectively.}
\label{fig:RT-only}
\end{figure}

\setlength{\tabcolsep}{2.2pt}
\begin{table}
\centering
\caption{Table of physical parameters determined from the photometric data. The columns from left to right list the Oo group, and the columns 2 and 3 list the medians of the pulsation periods and the mean metallicities for the Oosterhoff groups. Columns 4 and 5 list the median effective temperatures and the rise time (RT).}
\label{tsb:physparamOo}
\begin{tabular}{lccccccc}
\hline
      & $P$\,[day] & [Fe/H]$_{\rm JK}$ & $T_{\text{eff}}$\,[K] & RT \\ \hline
Oo I  & 0.536  &  -1.04   & 6522 & 0.153 \\
Oo II & 0.674  &  -1.15   & 6100 & 0.178 \\ \hline
\end{tabular}
\end{table} 
We see a larger difference in pulsation periods than in GCs \citep[see tab.~3.2 in][]{Smith1995} which broadcasts the difference in metallicity. The difference in the RT, on the other hand, shows that light curves of Oo\,II stars are less skewed than those of Oo\,I stars. This effect is also observed in Milky Way GCs \citep[see fig.~10 in][]{Sandage-K-S-1981}. When we look at the distribution of the RT (see Fig.~\ref{fig:RT-only}) we see that stars with the shortest RT are solely associated with the Oo\,I group. These stars have the highest amplitudes and shortest pulsation periods. The vast majority of them belong to the HASP (high amplitude short period) group identified by \citet{Fiorentino2015}. 

\begin{figure} 
\includegraphics[width=\columnwidth]{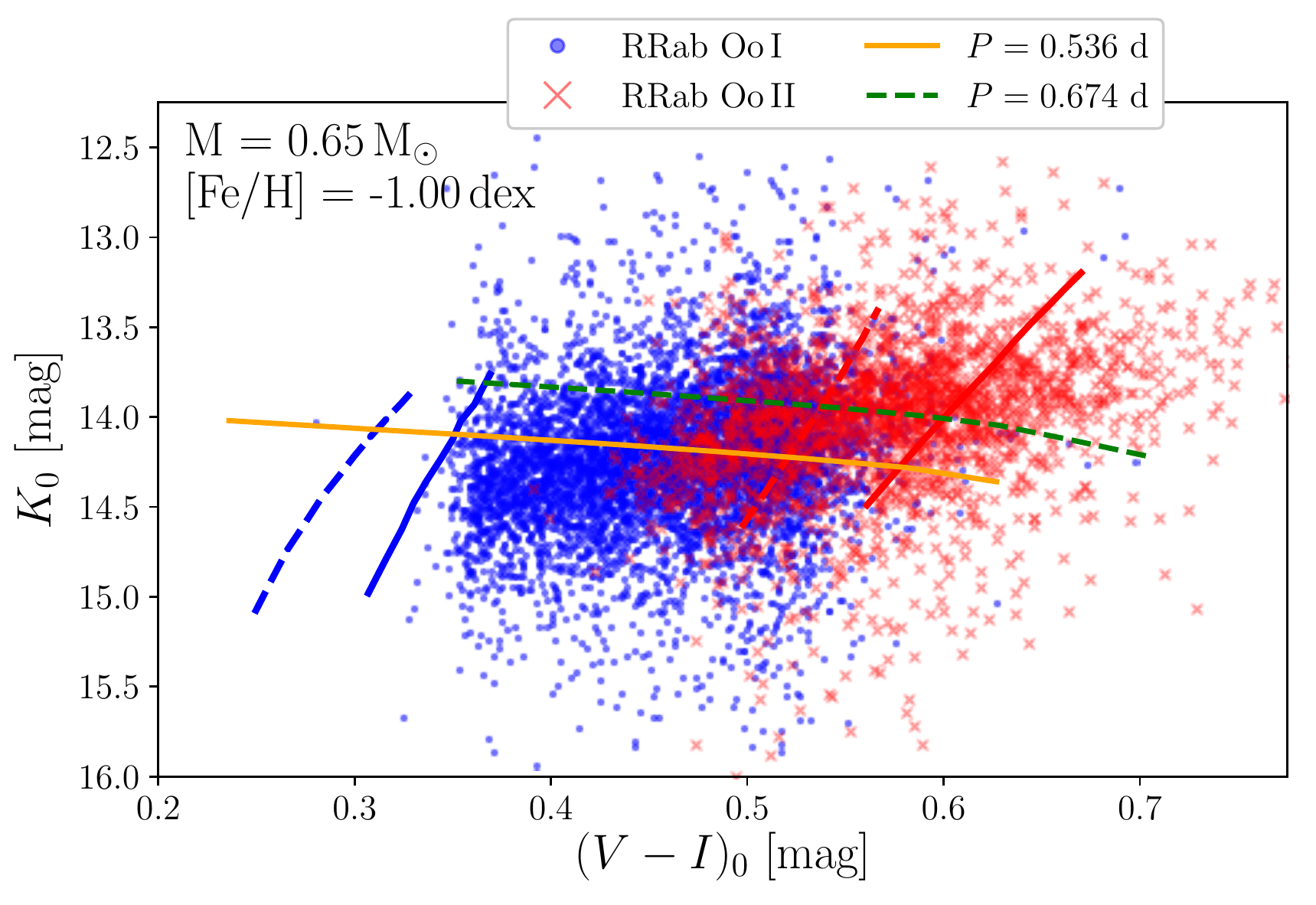}
\includegraphics[width=\columnwidth]{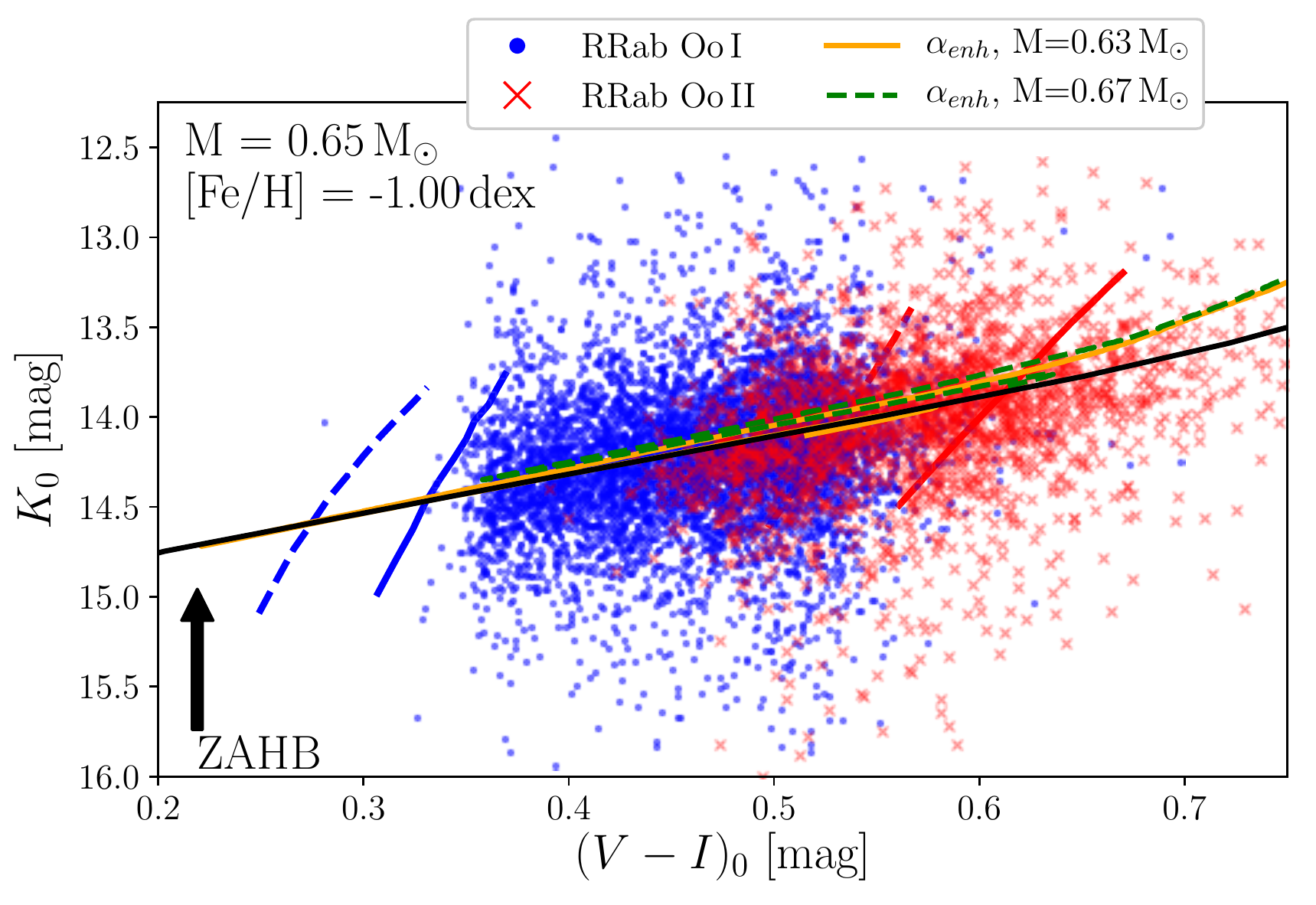}
\caption{The color-magnitude diagrams for the Oosterhoff groups from our sample stars compared with stellar pulsation models (top panel) and stellar evolution models (bottom panel). In the top and bottom panels red and blue lines stand for the red and blue edges of the instability strip for fundamental mode pulsators (solid lines) and the first overtones (dashed lines). In the top and bottom plots, blue dots and red crosses represent Oo\,I and Oo\,II variables, respectively. The green and orange lines in the top panel represent lines of constant period for the analyzed Oosterhoff groups, calculated using stellar pulsation models. The bottom panel depicts sample stars with horizontal branch models, where orange and green lines represent evolutionary tracks form masses 0.63\,M$_\odot$ and 0.67\,M$_\odot$ for Oo\,I and Oo\,II, respectively.}
\label{fig:CMD-OO-RT}
\end{figure}

The top and bottom panels of Fig.~\ref{fig:CMD-OO-RT} show color-magnitude diagrams (CMDs) of the stars in our sample. We compared the Oosterhoff groups with linear pulsating models (top panel) and stellar evolutionary models (bottom panel). We computed the boundaries of the instability strip (IS) using linear pulsation models \citep{Smolec2008}, calculated on a grid of the following physical parameters:
\begin{itemize}
\item Masses - $0.5 < M_\odot < 0.8$
\item Luminosities - $30 < L_\odot < 80$
\item Temperatures - $5500\,K < T_{\rm eff} < 8000\,K$
\end{itemize}
We used OPAL opacities \citep{Iglesias1996} and the solar heavy element mixture according to \citet{Asplund2009}. With the linear pulsation models we also calculated lines of constant period (yellow and green lines) for the median values of the pulsation periods in both Oosterhoff groups (see Table~\ref{tsb:physparamOo}). We computed the lines of constant period for the whole grid of stellar pulsation models for masses ranging from 0.5 to 0.8\,$M_\odot$ and searched for a case where we find a similar amount of stars above and below the lines of constant period in the CMD. The lines of constant period best describe our studied Oo\,I and Oo\,II populations for masses above 0.6\,$M_\odot$ (0.63\,$M_\odot$ for Oo\,I and 0.67\,$M_\odot$ for Oo\,II). We note that for this estimate we considered only stars with distances in the range from 6.4 to 10.3\,kpc (see Sec.~\ref{subsec:Kosty-range}), thus only the stars located in the approximate volume of the Galactic bulge. Therefore, we assumed these masses also for further comparison with stellar evolutionary models. 

We used the pre-computed horizontal branch tracks from the BaSTI database \citep{Pietrinferni2004}. We selected tracks for each Oosterhoff group separately. For the Oosterhoff type I stars we used horizontal branch tracks with the following parameters: $Z=0.002$, $M=0.63$\,$M_\odot$ and $\alpha$-enhanced abundances ([$\alpha$/Fe] = 0.4\,dex). For the Oosterhoff type II stars we used a different set of parameters $Z=0.001$, $M=0.67$\,$M_\odot$ and $\alpha$-enhanced abundances ([$\alpha$/Fe] = 0.4\,dex) due to their different median metallicity and mass from the models of stellar pulsation. We note that selecting $\alpha$-enhanced or scaled solar models does not change the results significantly. Thus, Oo\,II variables seem to have either similar or higher masses than their Oo\,I counterparts. A similar difference in masses between both Oosterhoff populations has been suggested in the past \citep{Catelan1992,Cacciari1993,Sandage2006}. We note that for pulsation and evolutionary models the $K_{0}$ and $\left(V-I \right)_{0}$ magnitudes were computed using static atmosphere models \citep{Kurucz2005}. 

Our results are in agreement with a study performed by \citet{McNamara2014} on GCs, where they demonstrated that Oo\,II variables have larger radii and are brighter in comparison with the Oo\,I stars. They also found 4 pulsators associated with the Oo\,III group. Overall we see that the Oo\,II group is on average brighter (see the top and bottom panel of Fig.~\ref{fig:CMD-OO-RT}) by approximately 0.23\,mag in the dereddened $K_{s}$ band. Moreover, the mean magnitudes of stars from this group are redder (by $\left(V - I \right)_{0} \approx$ 0.1\,mag) than those of their Oo\,I counterparts. Furthermore, the masses of Oo\,I and II seem to be different as well. At this point, it is necessary to emphasize that the estimated physical parameters are strongly correlated with the pulsation periods and metallicities. Therefore, the difference between the Oo\,I and II populations in both parameters will propagate into the others as well.

\section{Spatial distribution of the Oosterhoff groups} \label{sec:SpatDist}

In this section, we analyze the spatial distribution of both Oosterhoff groups in the Galactic bulge, based on their coordinates and calculated distances. We search for possible overdensities within the groups, differences in their positions in the bulge, and analyse their positions with respect to the Galactic bar.

\subsection{Oosterhoff variables in Galactic coordinates} \label{subsec:Kosty-range}

For this part we used the known Galactic coordinates and distances of our sample stars. First, the distances to individual stars were projected on the Galactic plane by multiplying them with a factor of cos\,$b$. Then we defined three regions: \textit{foreground}, \textit{center}, and \textit{background}. As the \textit{foreground} of the Galactic bulge we selected stars with $d \leq 6.4\,\rm kpc$ and for \textit{background} variables we applied a $d \geq 10.3\,\rm kpc$ condition. These cuts were selected based on the spatial density of our sample, to median distance to the Galactic bulge of 8.3\,kpc, assuming a bulge radius of approximately 2\,kpc (more than 80\,\% of our sample confined within selected radius, see Fig.~\ref{fig:densityProfile}, \ref{fig:two-point}). Subsequently, we divided the central region into 1287 cuboids with a edge length of 1.5\,deg edge and a length of 0.3\,kpc, and calculated the ratio of Oo\,I and Oo\,II variables in each frustum. The ranges into which we divided the central region were as follows: $l= \left( -8 ; 8.5\right)\rm\,deg$, $b= \left( -7.5 ; 6.0\right)\rm\,deg$ and $d= \left( 6.4 ; 10.3\right)\rm\,kpc$.

\begin{figure*} 
\includegraphics[scale=.1214]{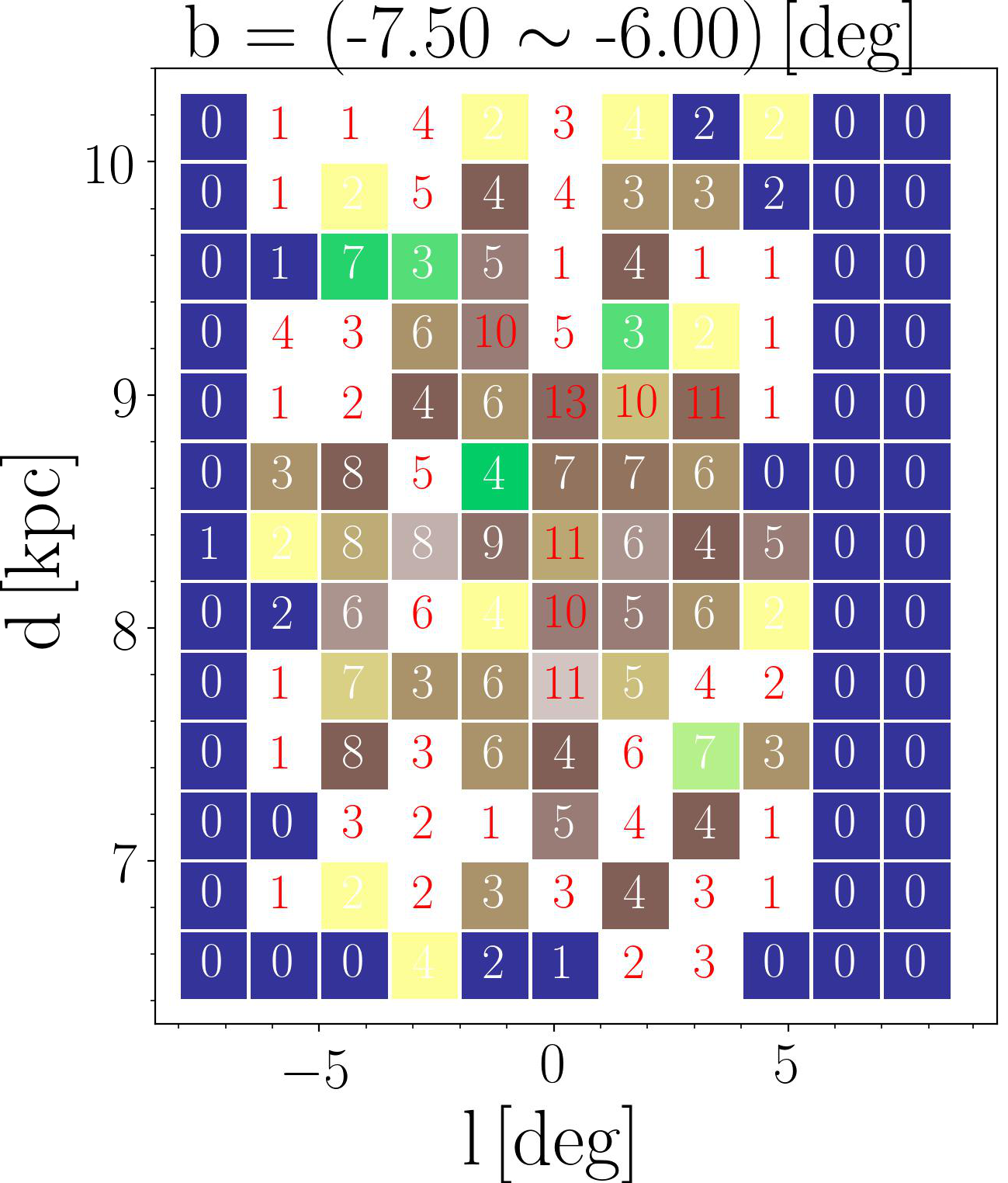}
\includegraphics[scale=.1214]{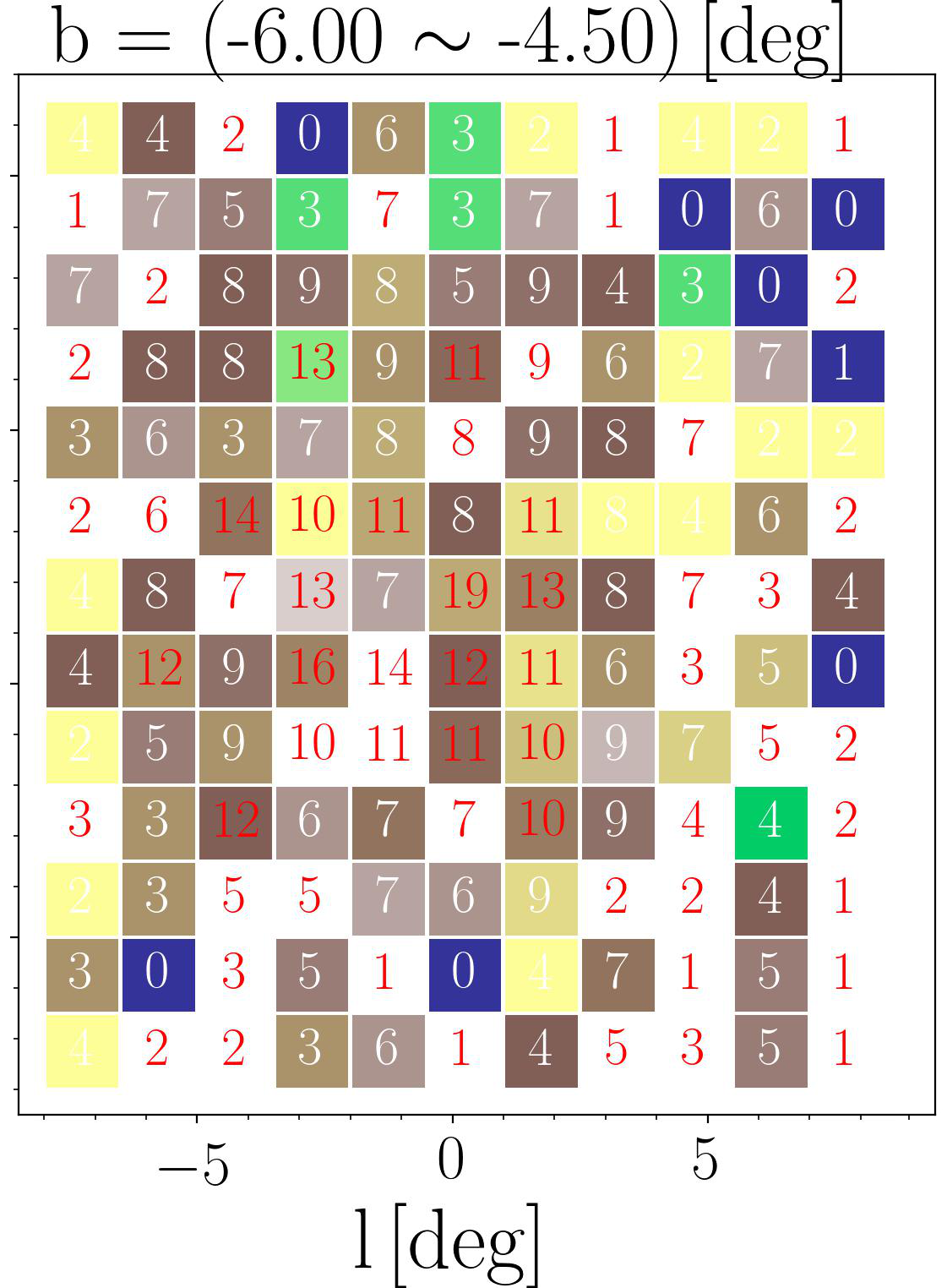}
\includegraphics[scale=.1214]{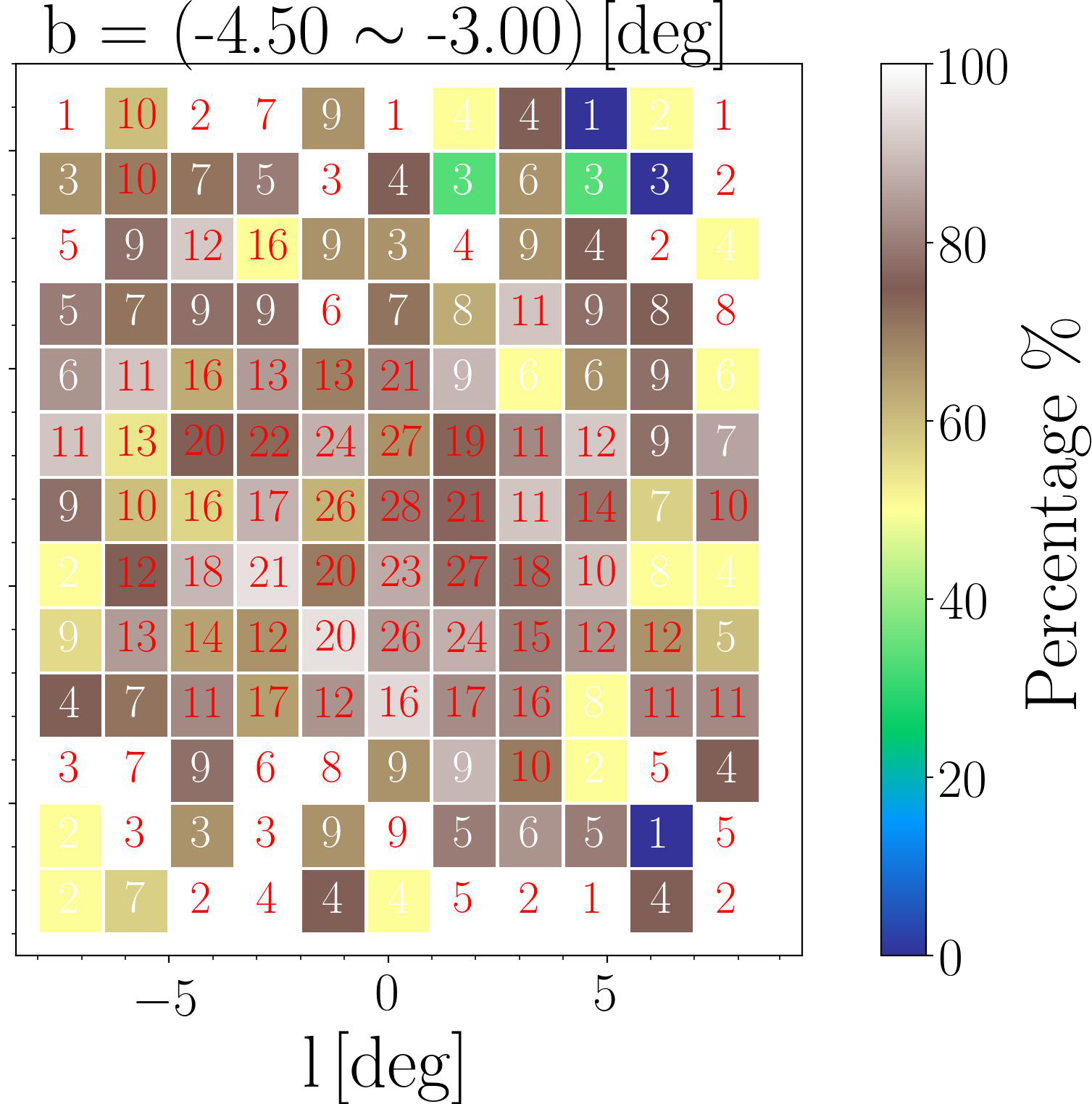} \\ \vspace{0.4cm}
\includegraphics[scale=.1214]{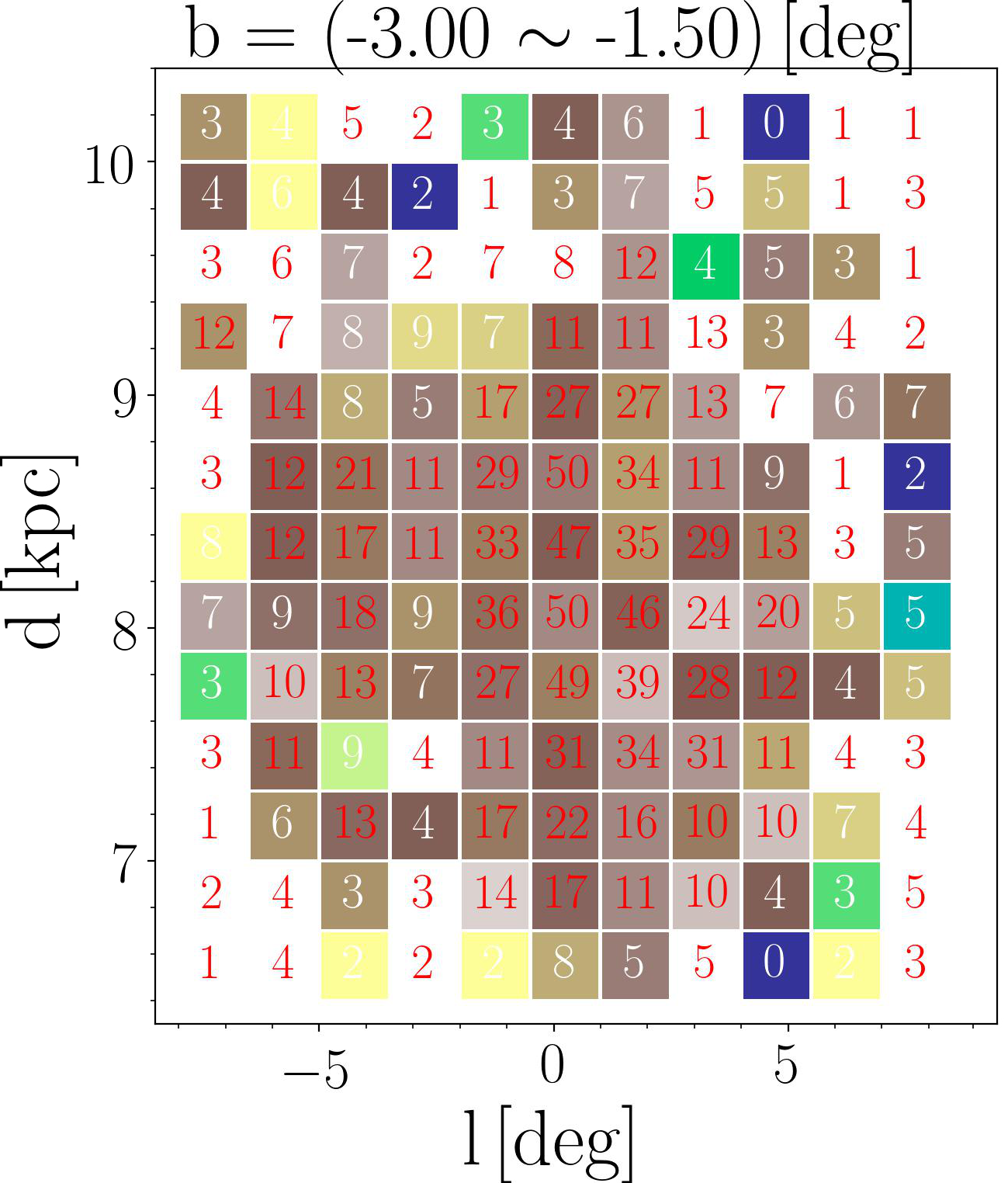}
\includegraphics[scale=.1214]{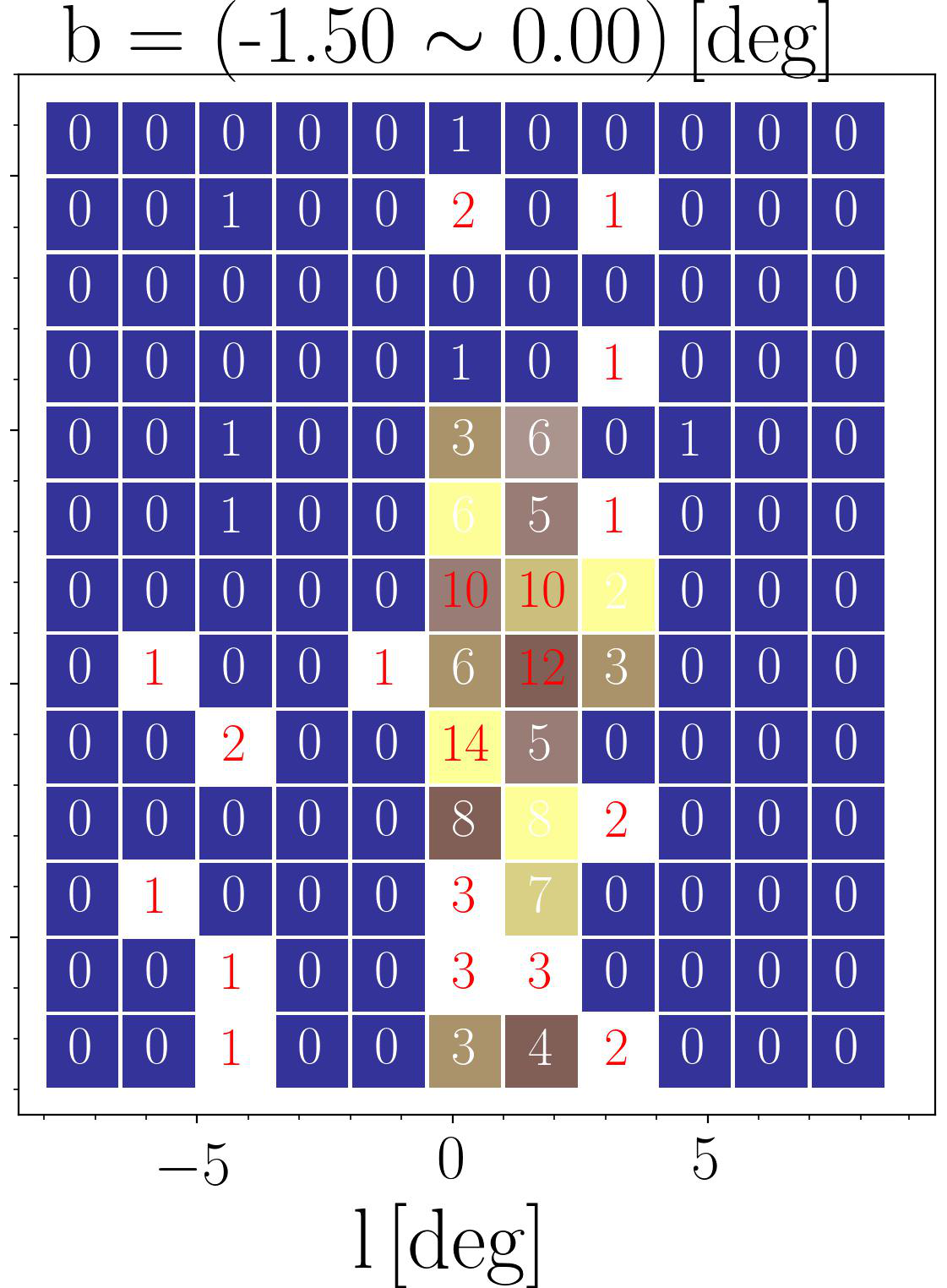}
\includegraphics[scale=.1214]{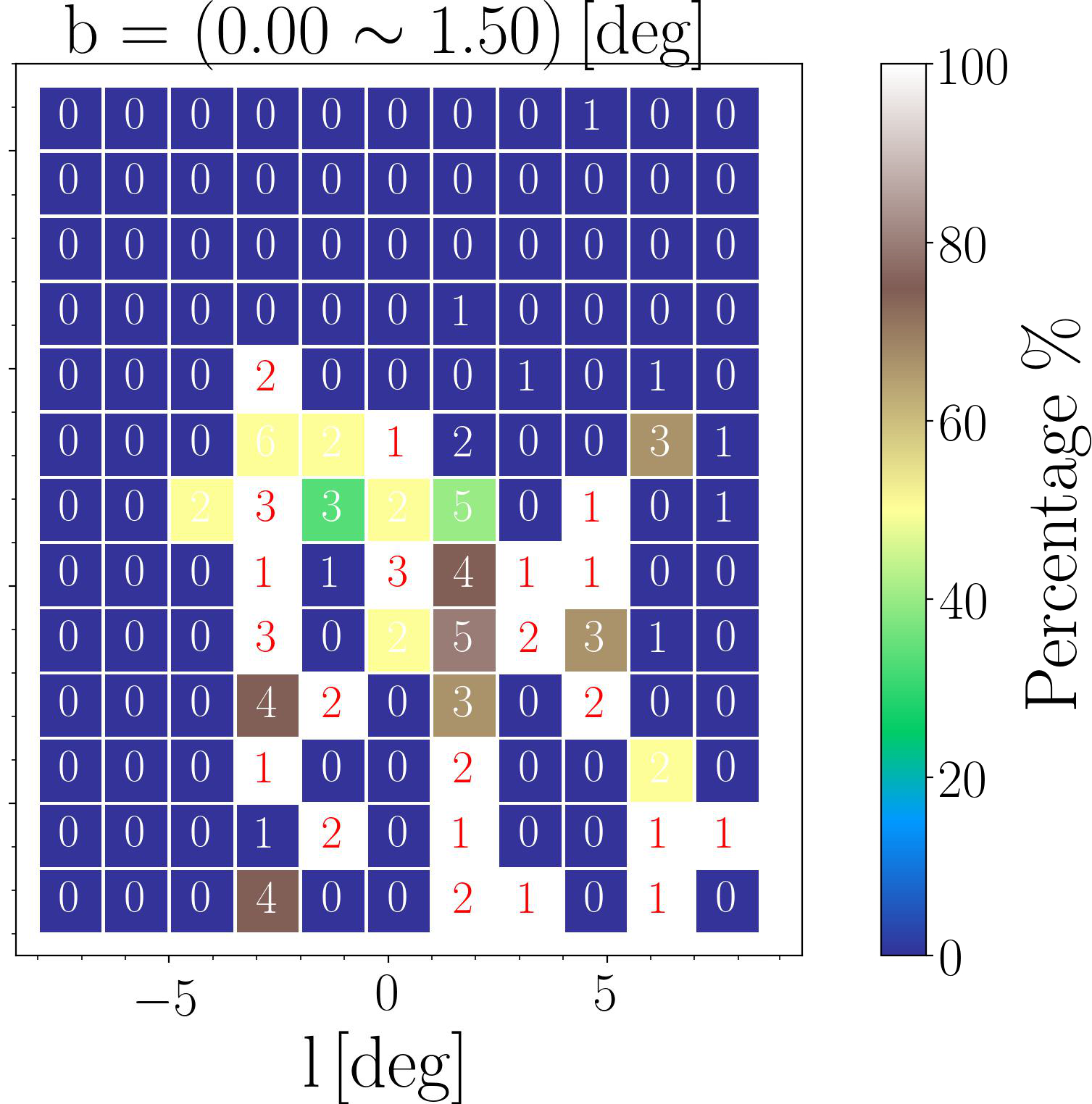} \\ \vspace{0.4cm}
\includegraphics[scale=.1214]{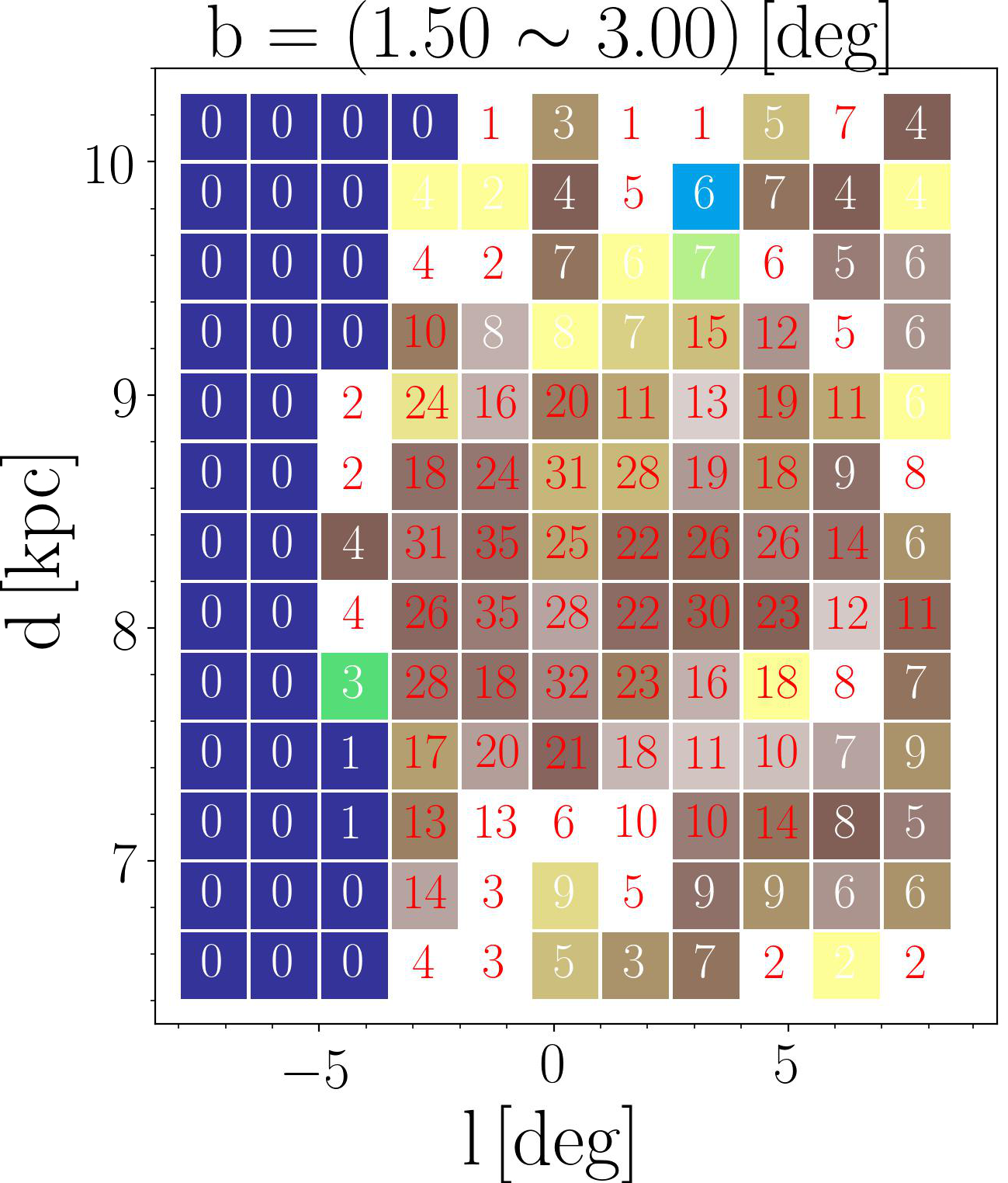}
\includegraphics[scale=.1214]{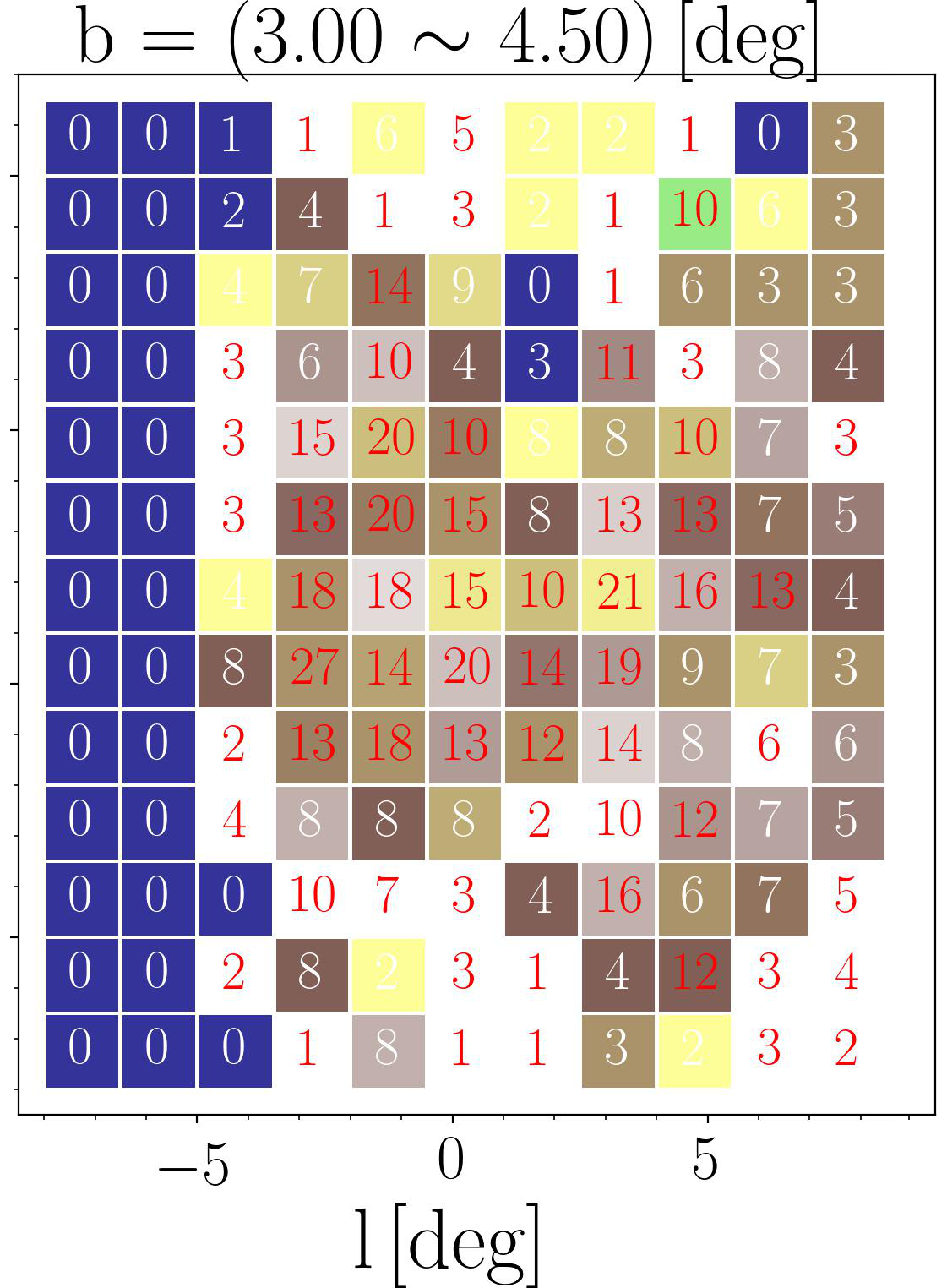}
\includegraphics[scale=.1214]{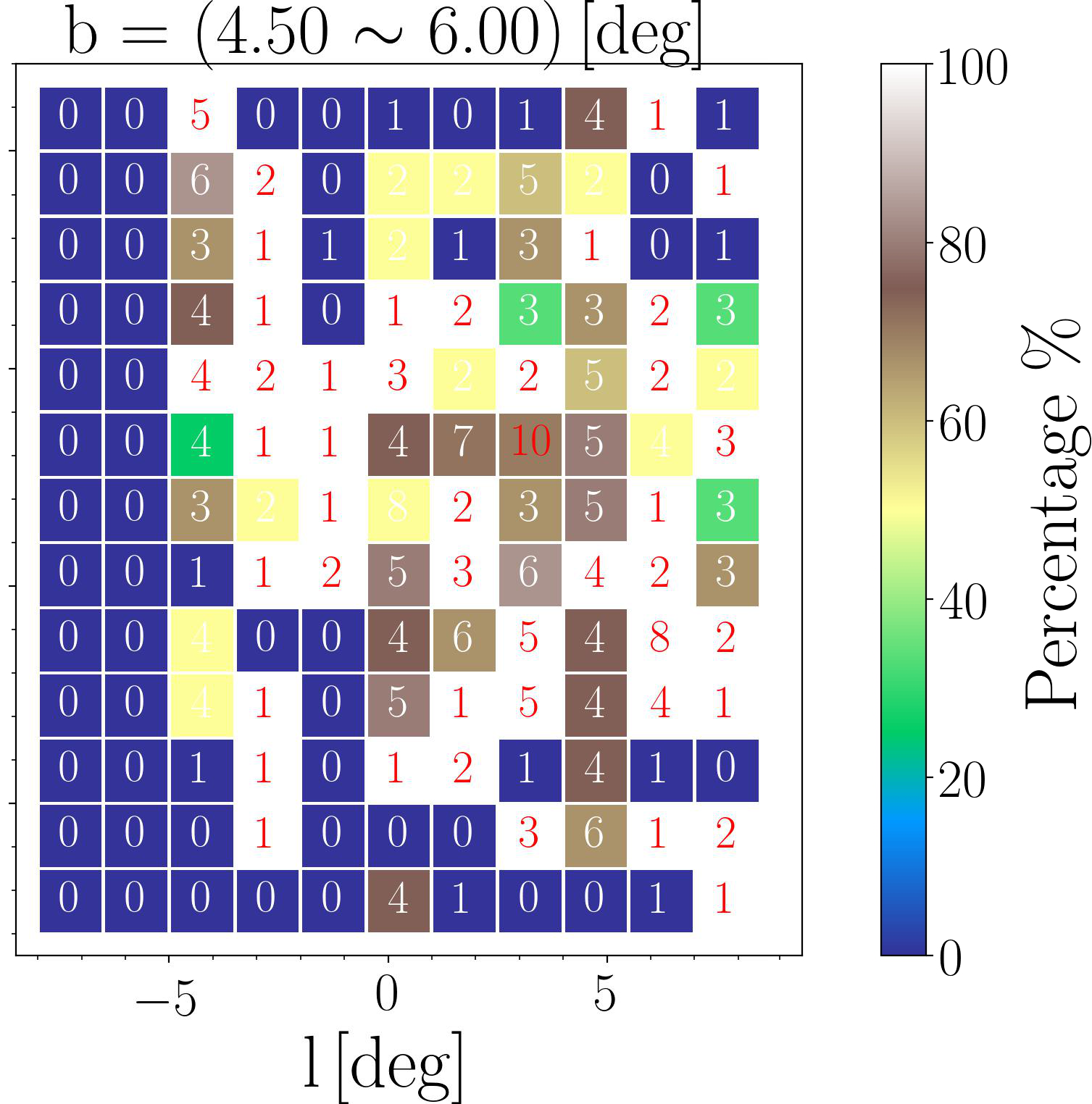} \\
\caption{Color maps for the spatial distribution of the studied RRab pulsators based on the fraction of the Oo\,I group using distance vs. Galactic longitude in selected stripes in Galactic latitude coordinates. From the top-left to the bottom-right panels we displayed slices going from $-6.75$ up to $5.25$\,deg with $1.5$\,deg steps. The white/brown colors mark the regions with the majority of Oo\,I variables, and green/yellow regions represent the regions with equivalent or higher number of Oo\,II variables. The blue regions denote in most cases the sectors with the lowest number of stars or those where the ratio between the Oosterhoff groups is more in favor of the Oo\,II group. These regions were in most cases not observed by the OGLE team. The numbers in each grid represent the total number of RR~Lyraes that were found in that region.}
\label{x-y-z-OO-density-galactic-coordinates}
\end{figure*}

Then, we divided our cuboid sample into stripes based on the Galactic latitude creating nine slices, each composed of 143 cuboids. The fraction of Oo\,I group stars in each slice is shown in Fig.~\ref{x-y-z-OO-density-galactic-coordinates}. This figure shows in nine panels color maps of the fraction of Oo\,I stars in different slices of the central region of the Galactic bulge. Overall, we do not see any large-scale structures traced by an excess of Oo\,I variables as compared to the characteristic ratio between Oo\,I and Oo\,II for the bulge as a whole (25\,\%; see Sec.~\ref{SepOogroupsSubSec}). We see some small structures but when closely inspected they do not show any possible overdensity in one or the other Oosterhoff type.

In Table~\ref{tab:spatDist-z-x-y}, we listed median values in $d$ for both Oosterhoff groups in individual regions, together with their fraction in each segment. From this table we see that in almost all studied regions the median values of $d$ are higher for Oo\,II stars. It is necessary to add that within the absolute-median-deviation for Oo\,I and II ($\pm 0.6$\,kpc), differences in distances are not significant. The difference in the representation of both Oosterhoff groups is the most striking between \textit{foreground} and \textit{background}. In the \textit{foreground} more than 80\,\% variables are Oo\,I stars, while in the \textit{background} their numbers go down to 63\,\%. For the central region the representation of Oo\,II increases up to 30\,\%. The discrepancy between \textit{foreground}, \textit{background} and central regions might also be an incompleteness effect, since the Oo\,II stars are on average brighter and thus easier to detect at larger distances or in high-extinction regions.

\setlength{\tabcolsep}{7pt}
\begin{table}
\centering
\caption{Table for the median values of the distance in each selected segment with respect to the Oosterhoff group. Column 1 lists the studied regions, columns 2 and 3 and columns 4 and 5 mark the median values of the distance and fraction of a given Oosterhoff group, respectively.}
\label{tab:spatDist-z-x-y}
\begin{tabular}{lcccc} \hline
 & \multicolumn{2}{c}{Oo I} & \multicolumn{2}{c}{Oo II} \\ 
       Region           & d\,[kpc]            & \%        & d\,[kpc]            & \%         \\ \hline
\textit{foreground}$_{\left(d \leq 6.4\,\rm kpc \right) }$        & 5.85 & 81.4 & 5.94 & 18.6 \\
$b = \left \langle -6.75 \pm 0.75 \right \rangle$\,[deg]             & 8.35 & 73.5 & 8.50 & 26.5 \\
$b = \left \langle -5.25 \pm 0.75 \right \rangle$\,[deg]             & 8.29 & 75.5 & 8.48 & 24.5 \\
$b = \left \langle -3.75 \pm 0.75 \right \rangle$\,[deg]             & 8.25 & 77.1 & 8.37 & 22.9 \\
$b = \left \langle -2.25 \pm 0.75 \right \rangle$\,[deg]             & 8.14 & 76.7 & 8.21 & 23.3 \\
$b = \left \langle -0.75 \pm 0.75 \right \rangle$\,[deg]            & 7.94 & 69.2 & 8.11 & 30.8 \\
$b = \left \langle 0.75 \pm 0.75 \right \rangle$\,[deg]             & 7.74 & 67.4 & 8.42 & 32.6 \\
$b = \left \langle 2.25 \pm 0.75 \right \rangle$\,[deg]              & 8.17 & 75.5 & 8.31 & 24.5 \\ 
$b = \left \langle 3.75 \pm 0.75 \right \rangle$\,[deg]              & 8.20 & 75.4 & 8.43 & 24.6 \\ 
$b = \left \langle 5.25 \pm 0.75 \right \rangle$\,[deg]             & 8.31 & 73.8 & 8.57 & 26.2 \\
\textit{background}$_{\left(d \geq 10.3\,\rm kpc \right) }$        & 11.36 & 63.1 & 11.58 & 36.9 \\  \hline
\end{tabular}
\end{table}

To conclude, we do not see any major structures in the Galactic bulge formed by either of the Oosterhoff groups. We do see some small iregularities in their distributions, mainly in the front and back part of the Galactic bulge. The differences in number fraction of \textit{foreground} and \textit{background} regions might be connected to an observational bias, since the Oo\,I and II populations differ in luminosity.

\subsection{Density profile of bulge RR~Lyraes and the Galactic bar}

\begin{figure} 
\includegraphics[width=\columnwidth]{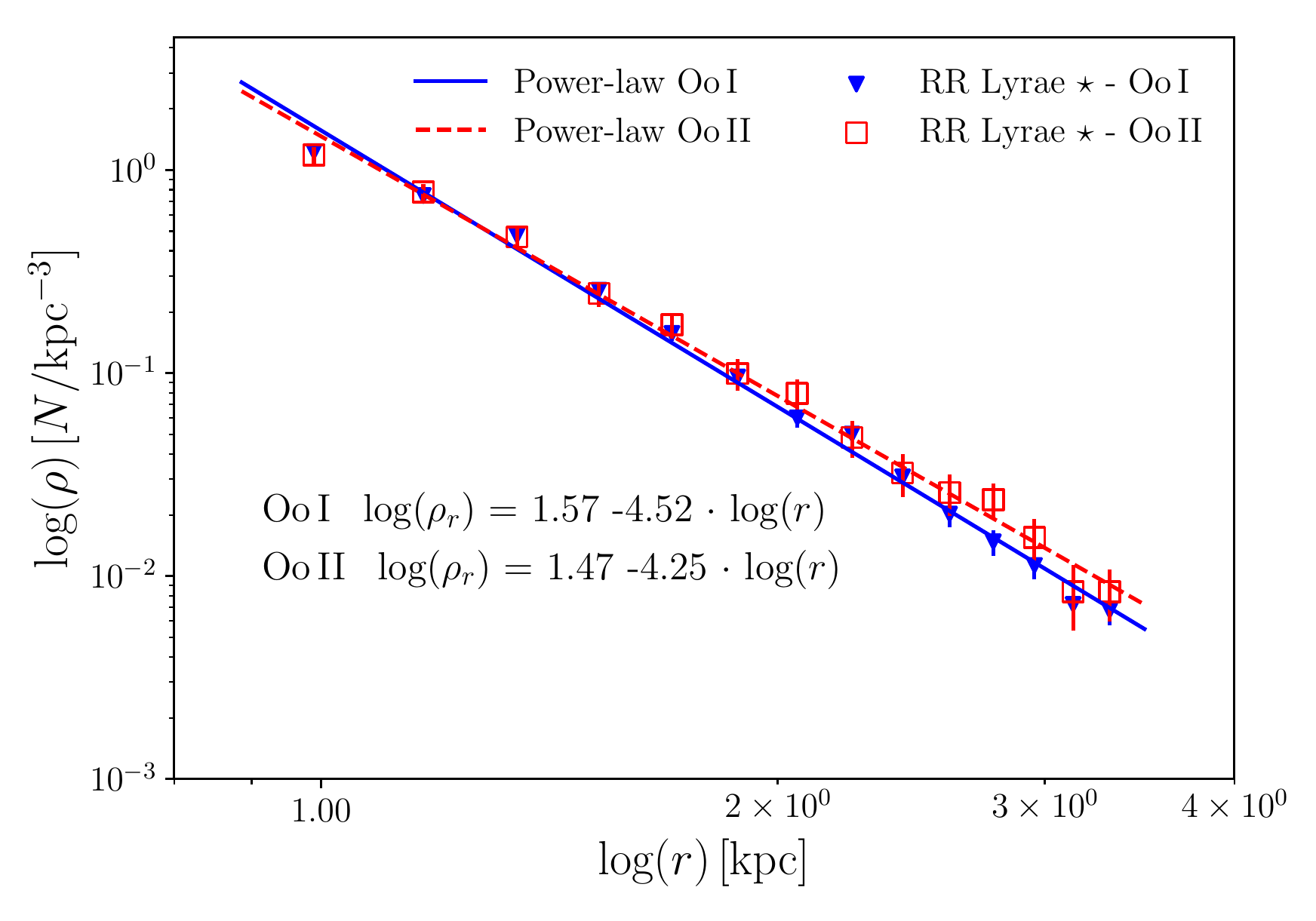}
\caption{The density profile in the studied Oosterhoff groups in the Galactic bulge. The panel shows the density profile of RR Lyraes in the Oo\,I and II populations assuming a spherical model of the Galactic bulge. The blue dots and red squares represent Oo\,I and Oo\,II stars, respectively. The errors in the individual bins were calculated using a Monte Carlo simulation.}
\label{fig:densityProfile}
\end{figure}

For the construction of the density profile we transformed Galactic coordinates and distances of individual stars into a Galactocentric Cartesian coordinate system using the following equations:
\small{
\begin{equation} 
x = \left(- d + d_{\rm cen}\right)  \cdot \left ( \text{cos}\,b \cdot \text{cos} \,b_{\rm cen} \cdot \text{cos} \left ( l - l_{\rm cen} \right )+ \text{sin}\,b \cdot \text{sin}\,b_{\rm cen} \right),
\end{equation}
\begin{equation} 
y = d \cdot  \text{cos}\,b \cdot \text{sin} \left ( l - l_{\rm cen} \right ),
\end{equation}
\begin{equation} 
\begin{split}
z = d \cdot  \left ( \text{sin}\,b \cdot \text{cos}\,b_{\rm cen}  \right ) - d \cdot \text{cos}\,b \cdot \text{sin}\,b_{\rm cen} \cdot \text{cos} \left ( l - l_{\rm cen} \right ),
\end{split}
\end{equation}}
\normalsize
\noindent where $b_{\rm cen}$ and $l_{\rm cen}$ denote the center of the Galactic bulge ($b_{\rm cen} = l_{\rm cen} = 0^{\circ}$), The $d_{\rm cen}$ stands for the median of the estimated distances for stars from our sample, which we will denote as distance to the Galactic center. In our transformation, the coordinate axes correspond to each other in the following way: $x \sim d$ (increases in positive direction towards the Galactic anticenter), $y \sim l$ (positive in the direction of Galactic rotation) and $z \sim b$ (positive in the direction to the Galactic North pole). Using the Cartesian coordinates we then calculated the radius $r$:
\begin{equation} \label{eq:simple_power_law}
r = \sqrt{x^{2} + y^{2} + z^{2}}.
\end{equation}
Subsequently, using the radius $r$ we divided our sample stars, with respect to each Oosterhoff group, into individual annuli with a width of roughly 0.18\,kpc except for the first bin, which had a larger width of 0.2\,kpc to account for the low number of pulsators in the central region. We cut off at a radius of $r = 3.4$\,kpc, due to the fact that for higher radii we would include only the stars behind the Galactic bulge at distances higher than 11.7\,kpc. Furthermore, we calculated the number density of stars in individual spherical shells around the Galactic bulge. All calculated densities were normalized using the density of the first bin. This bin was afterwards removed from the density plot due to possible incompleteness because of its position close to the Galactic center. The spherical distribution was fitted with a power-law in the following form:
\begin{equation} 
\text{log}\,(\rho_{r}) = k + n \cdot \text{log}\,(r),
\end{equation}
where $\rho_{r}$ stands for RR~Lyrae number density, and $k$ and $n$ are parameters of the model. Fig.~\ref{fig:densityProfile} depicts the resulting density profile for the RR~Lyrae distribution in the Galactic bulge. The spherical distribution for both Oosterhoff populations can be described by a simple power-law. 

Error bars for an individual bin in the density profile were calculated in the following way. We assumed that the distribution of log$(r)$ for both Oosterhoff groups follows a Gaussian distribution. Using average values and the standard deviation for log$(r)$ we randomly generated 50\,000 distributions for an individual Oosterhoff population and divided them into the same bins as for the selected sample stars. Subsequently, we calculated the standard deviation in individual bins, which is depicted as the aforementioned error bar in Fig.~\ref{fig:densityProfile}. Based on the calculated deviations in the bins, we detect at some radii a statistically significant difference between the Oo\,I and Oo\,II stellar density, especially in the outer regions.

From Fig.~\ref{fig:densityProfile} we see that in the central regions (around $r=1$\,kpc from the center) the normalized density of Oo\,II variables is higher than that of Oo\,I pulsators (although not statistically significant). This is potentially due to the fact that Oo\,II stars are on average brighter than their Oo\,I counterparts, thus it is easier to detect them at such large distances or in regions with higher extinction. As we move outward, the densities become comparable and Oosterhoff populations seem evenly distributed. In the regions with higher $r$ (approximately from $r = 2.0$\,kpc) the difference between the density of Oo\,II stars and their Oo\,I counterparts becomes equal and in some cases even higher and statistically significant, based on the calculated errors. This higher incidence rate of Oo\,II variables could be ascribed to a region behind the Galactic bulge (see Sec.~\ref{subsec:Kosty-range}). 

Here it is important to address the depth of the OGLE survey and its completeness. \citet{Udalski2015} estimated the depth of dense OGLE-IV fields to reach $\approx$\,20.5\,mag. They found OGLE-IV to be complete to $\approx$\,18.5\,mag in the $I$-band for individual sources in dense fields. The completeness of OGLE-IV RR~Lyraes depends on many aspects, e.g. the amplitude of changes and the brightness, the number of observations, the pulsation mode etc. For detection and correct classification of individual RR~Lyrae stars the above value might not be the true completeness of the OGLE-IV sample. \citet{Soszynski2014} used overlapping fields observed by OGLE-IV to evaluate completeness. For fundamental-mode RR~Lyrae pulsators they estimate the completeness to be above 95\,\% for variables brighter than $I=17$\,mag. 

\setlength{\tabcolsep}{5pt}
\begin{table}
\centering
\caption{Table for differences in density between the Oosterhoff groups in the Galactic bulge. Column 1 represents annuli in radius $r$ extending from the Galactic center. The columns 2 and 3 list the normalized stellar density for the Oo\,I and Oo\,II populations, respectively. Column 4 represents absolute differences between the densities of the Oosterhoff populations in individual bins and the uncertainty of this difference. The bold-faced rows mark regions where the difference is statistically significant.}
\label{tab:diff-Density-rozdily}
\begin{tabular}{lccc}
\hline 
$r$\,[kpc]     & $\rho_{\rm Oo\,I}$\,[N$_{\star}$/kpc$^{3}$] & $\rho_{\rm Oo\,II}$\,[N$_{\star}$/kpc$^{3}$] & $\vert \rho_{\rm Oo\,I} - \rho_{\rm Oo\,II}\vert$ \\ \hline 
0.99	&	1.2326	&	1.1851	&	0.0475	$\pm$	0.1263 \\
1.17	&	0.7480	&	0.7804	&	0.0324	$\pm$	0.0832 \\
1.35	&	0.4755	&	0.4678	&	0.0077	$\pm$	0.0559 \\
1.53	&	0.2559	&	0.2466	&	0.0092	$\pm$	0.0387 \\
1.70	&	0.1567	&	0.1728	&	0.0161	$\pm$	0.0275 \\
1.88	&	0.0955	&	0.0994	&	0.0038	$\pm$	0.0198 \\
2.06	&	0.0601	&	0.0795	&	\textbf{0.0194	$\pm$	0.0147} \\
2.24	&	0.0500	&	0.0481	&	0.0019	$\pm$	0.0110 \\
2.42	&	0.0308	&	0.0322	&	0.0014	$\pm$	0.0084 \\
2.60	&	0.0202	&	0.0258	&	0.0056	$\pm$	0.0065 \\
2.78	&	0.0147	&	0.0238	&	\textbf{0.0091	$\pm$	0.0051} \\
2.95	&	0.0113	&	0.0155	&	\textbf{0.0042	$\pm$	0.0040} \\
3.13	&	0.0072	&	0.0084	&	0.0011	$\pm$	0.0033 \\
3.31	&	0.0068	&	0.0084	&	0.0016	$\pm$	0.0026 \\ \hline

\end{tabular}
\end{table}

\subsection{Oosterhoff groups with respect to the Galactic bar}

In the past, several studies tried to use RR~Lyrae stars as tracers of the Galactic bar with different results \citep{Dekany2013,Pietrukowicz2015}. From the kinematical point of view, the kinematics of old population variables is inconsistent with the kinematics expected for a B/P bulge \citep{Kunder2016}.  

The position of the Galactic bar has been traced by intermediate-age stars \citep[red clump stars;][]{Nishiyama2005,Gonzalez2011,Wegg2013}. In order to explore a possible association with the Oosterhoff groups, we will use the bar positions found by \citet{Gonzalez2011}. They traced the bar above and below the Galactic center (Galactic latitudes $b = \pm 1$\,deg). To compare the distances of individual stars from the Galactic bar with respect to their Oosterhoff population we transformed coordinates and distances into the Cartesian coordinate system described above. We transformed the coordinates and distances from \citet{Gonzalez2011} for the Galactic bar as well. Subsequently, we calculated the Euclidian distances between individual pulsators and the Galactic bar.

\begin{figure*} 
\includegraphics[width=\columnwidth]{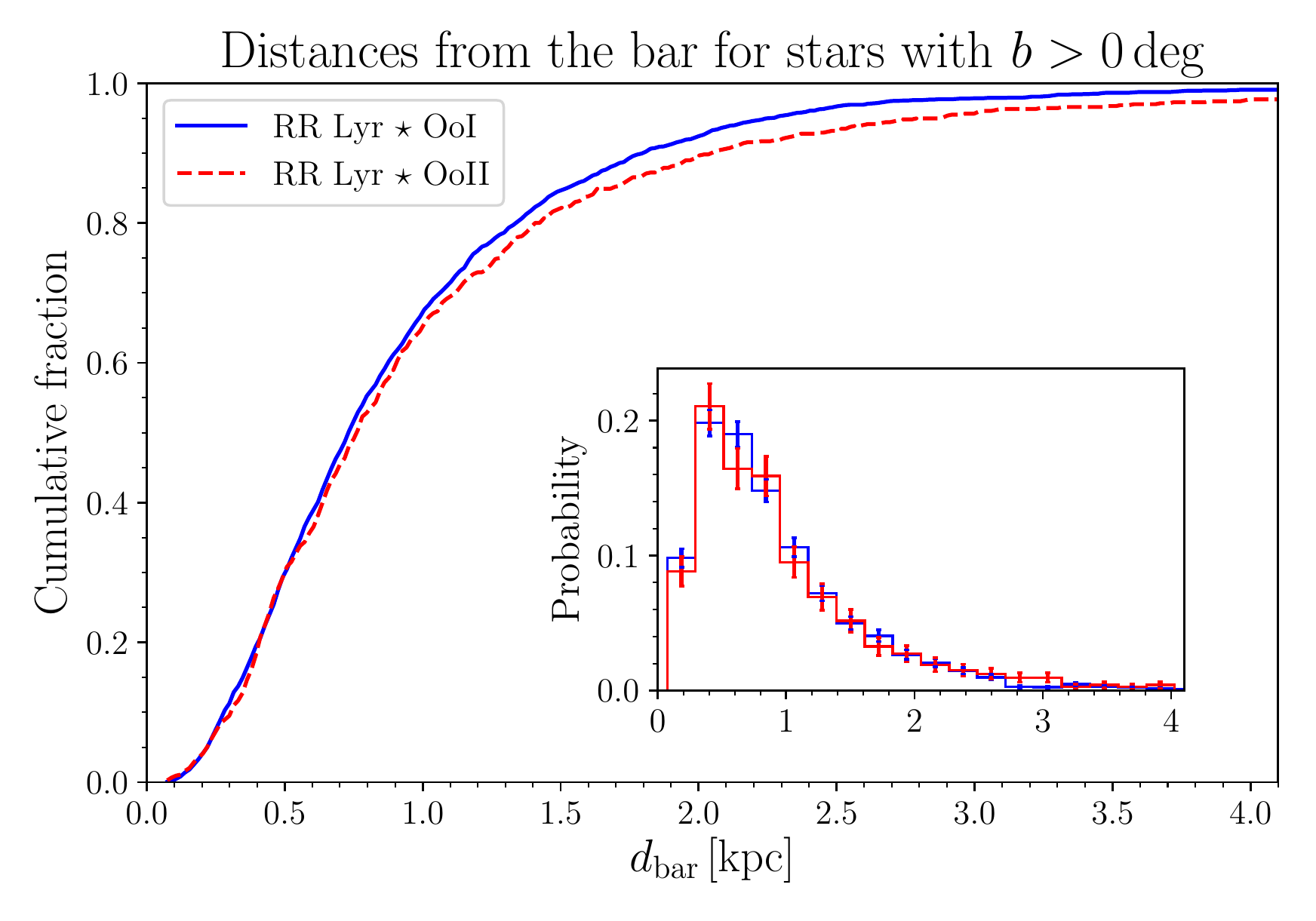}
\includegraphics[width=\columnwidth]{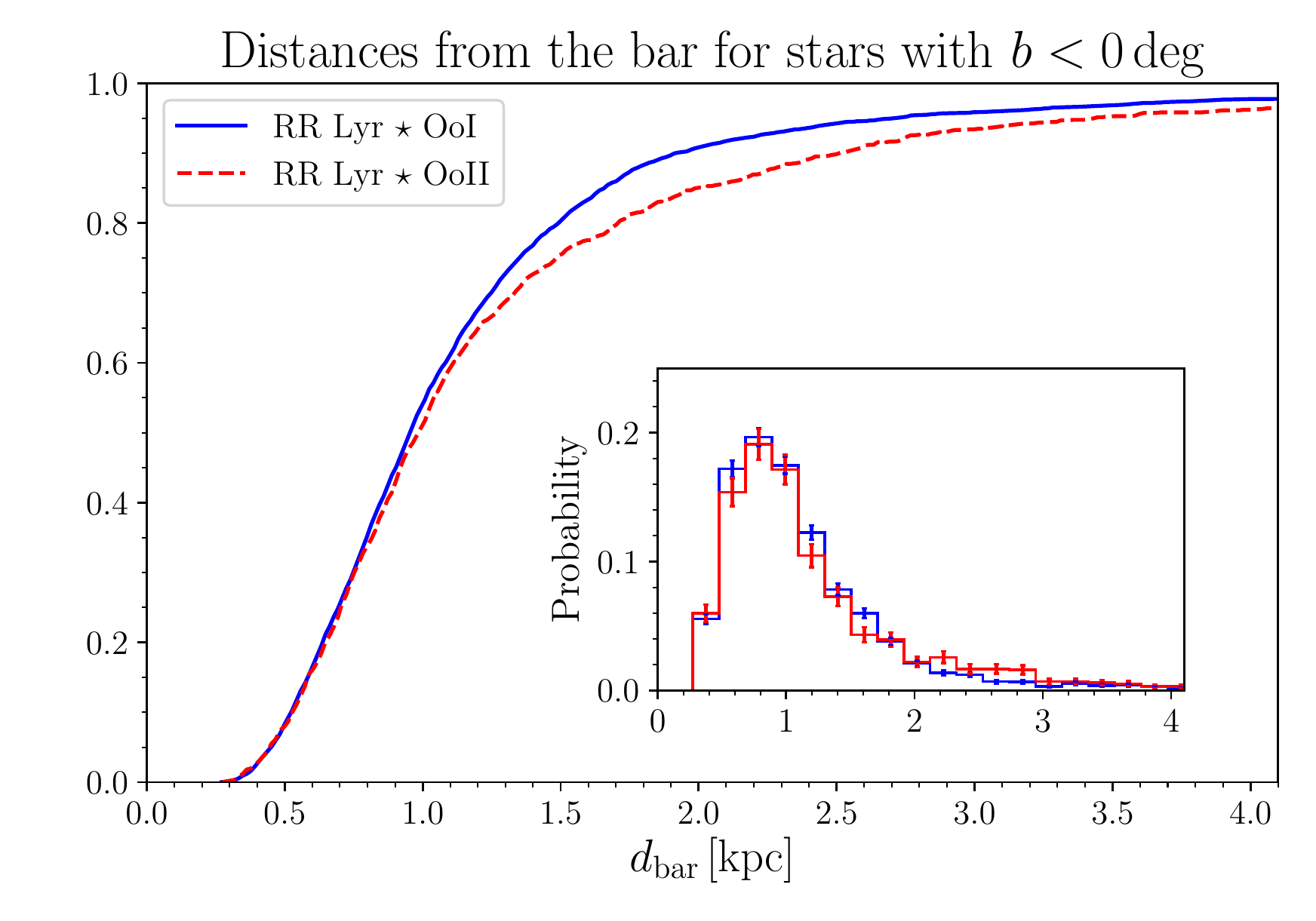}
\caption{The cumulative fractions and distributions (in the insets) of the distances from the bar for the studied Oosterhoff groups. The left-hand plot depicts the distribution above the Galactic plane where the red dashed lines stand for the Oosterhoff type II variables and the blue lines represent Oosterhoff type I pulsators. The right-hand plot represents a similar distribution but for stars below the Galactic plane with the same color-coding.}
\label{fig:BarDisty}
\end{figure*}

In Fig.~\ref{fig:BarDisty} we plot the cumulative fraction and normalized distributions of distances for our sample stars from the Galactic bar above (left-hand panel) and below (right-hand plot) the Galactic plane. We see that below and above the Galactic plane the Oo\,I stars peak at slightly shorter distances from the Galactic bar than the Oo\,II population. For Oo\,I stars the median values for distances from the Galactic bar are 0.75\,kpc and 0.97\,kpc above and below the Galactic plane, respectively. The Oo\,II component has a median distance of 0.77\,kpc and 1.00\,kpc above and below the Galactic plane, respectively. A similar effect is also seen in the inset of Fig.~\ref{fig:BarDisty} where we see that for distances close to the Galactic bar, the probability of finding Oo\,I stars is higher than for Oo\,II variables. Once we move further away from the bar (approximately at 1.15\,kpc away in radial direction) the probability of finding one or the other group becomes almost the same or higher for the Oo\,II group.

\begin{figure*} 
\includegraphics[width=\columnwidth]{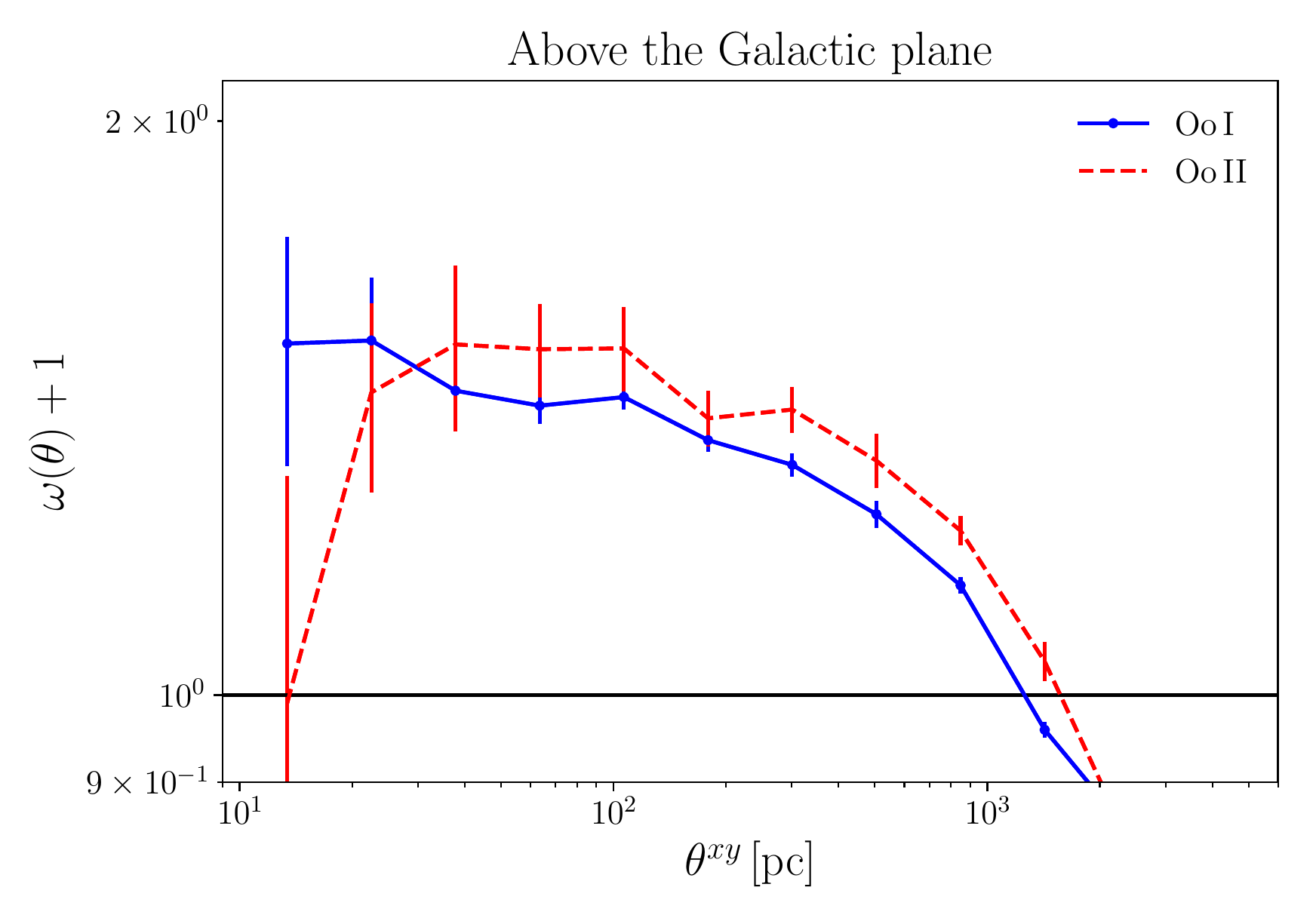}
\includegraphics[width=\columnwidth]{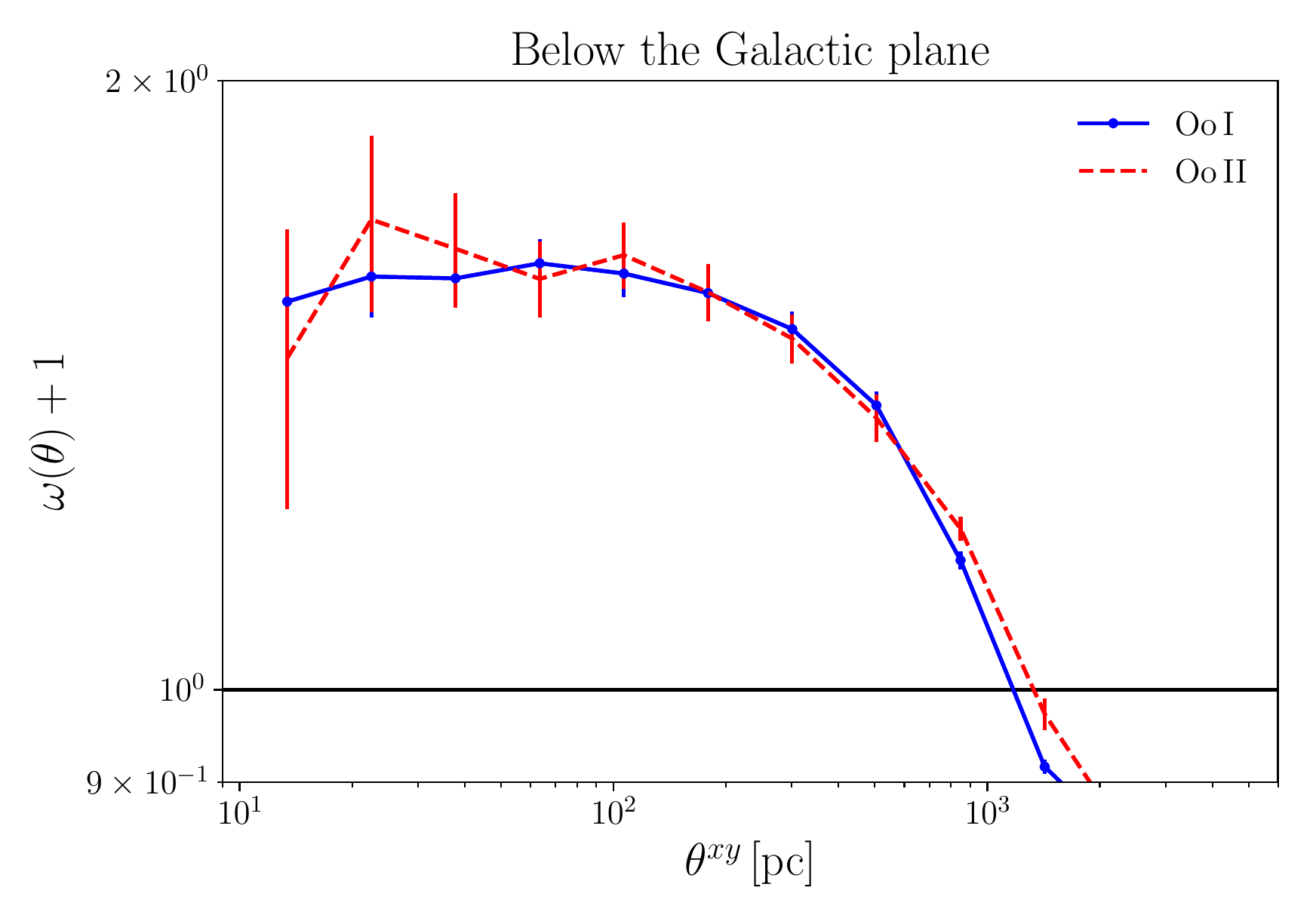} \\
\includegraphics[width=\columnwidth]{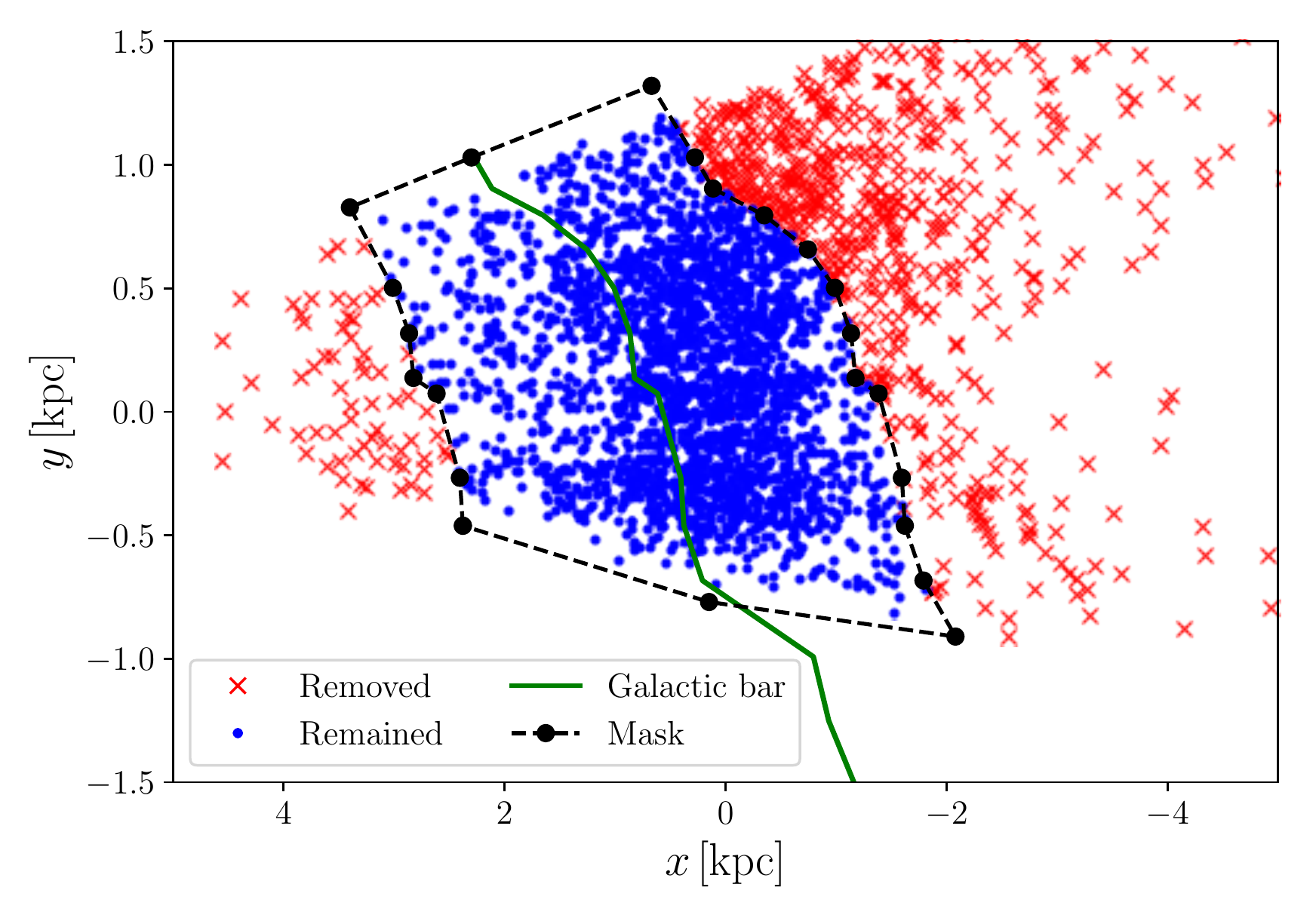}
\includegraphics[width=\columnwidth]{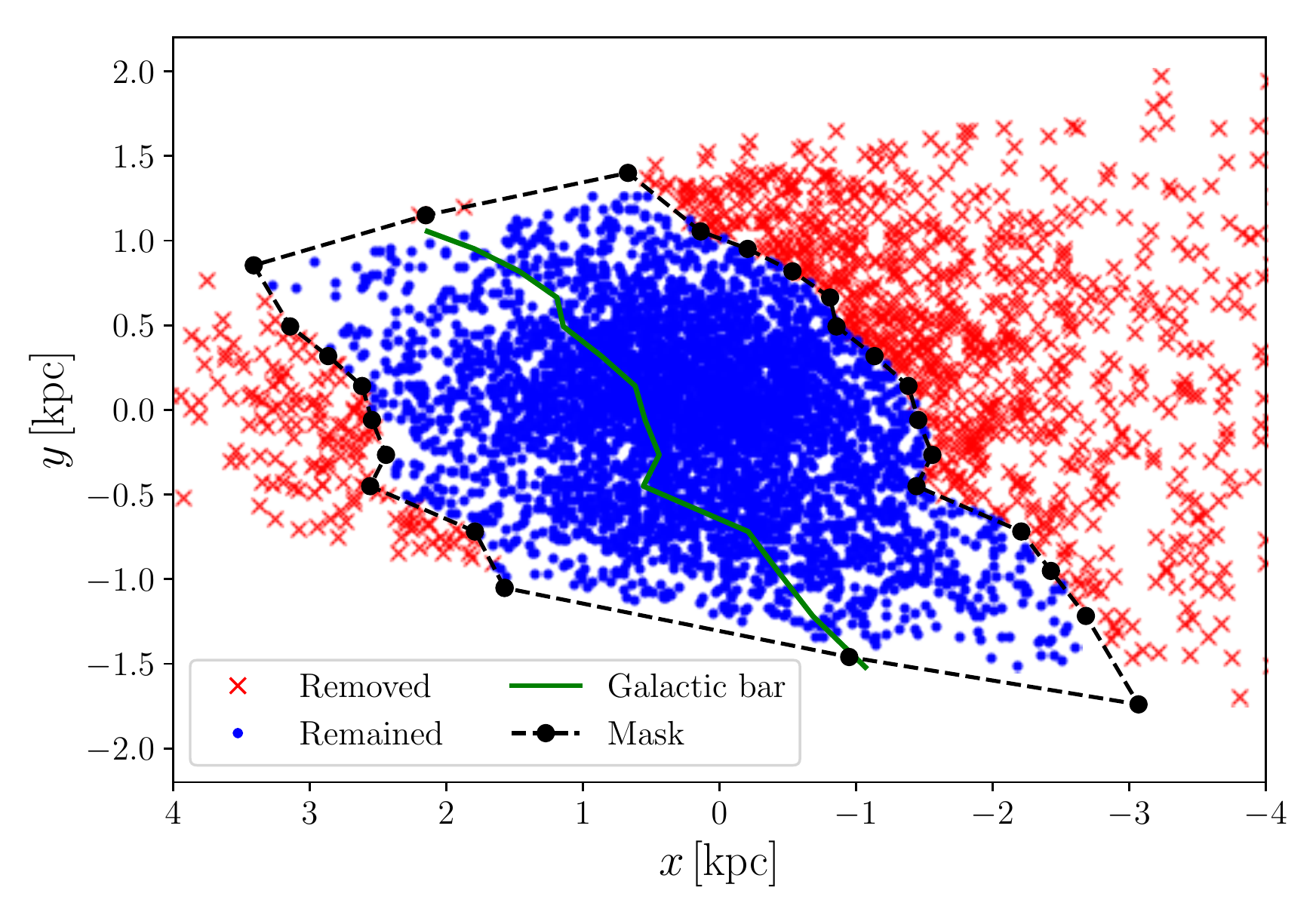} 
\caption{Two-point correlation function for the Oo\,I and II groups in the Galactic bulge. The two top panels show results of the two-point correlation function, $\omega(\theta)$, with blue and red lines denoting Oo\,I and Oo\,II stars, respectively, while $\theta$ represents the distance separation between the data and randomly generated points. The left-hand panel shows results for the region above the Galactic plane and the right-hand panel illustrates the region below the Galactic plane. The two bottom figures display masked regions (black dashed lines) with the Galactic bar (green solid line). The blue points stand for stars within the masked region, while red crosses lie outside the outline boundary.}
\label{fig:two-point}
\end{figure*}

To further study the possible correlation between the Galactic bar and RR~Lyrae stars we computed two-point correlation functions in order to quantify the clustering of stars in the vicinity of the Galactic bar. In the two-point correlation function $\omega(\theta)$ we quantify how likely it is to find two stars separated by a distance $\theta$ in a non-clustered randomly generated distribution \citep{Peebles1980}. Over the years several estimators have been introduced by e.g., \citet{Davis1983} and \citet{Hamilton1993}. In this study we will use the estimator from \citet{LS1993}, which reads as follows:

\begin{equation}
\omega(\theta) = \frac{r\cdot\left ( r-1 \right )}{n \cdot \left ( n-1 \right )} \frac{DD}{RR} - \frac{\left ( r-1 \right )}{n} \frac{DR}{RR} + 1,
\end{equation}
where the variables $r$ and $n$ represent total number of random points and of data points. The $DD$, $DR$, and $RR$ acronyms are pair counts of stars in the data catalog, data and random catalog, and random catalog, respectively. We selected regions around the Galactic bar in $x$ and $y$ coordinates above and below the Galactic plane with a range of $\pm$\,2\,kpc in $x$ direction for the position of the Galactic bar and removed stars lying outside the set borders (see bottom panels of Fig.~\ref{fig:two-point}). Two random catalogs were generated and masked in the same manner. 

The top panels of Fig.~\ref{fig:two-point} show the resulting two-point correlation functions for both Oosterhoff groups above and below the Galactic plane with respect to the Galactic bar. For the region below the Galactic plane we see that Oo\,I and II stars, in the vicinity of the Galactic bar, cluster in a similar manner. Neither of the groups seems to be more clustered. On the other hand, above the Galactic plane we see that the Oo\,II group seems to be more clustered in comparison with the Oo\,I component. However, results from the Kolmogorov-Smirnov test suggests that the Oo\,I and II groups are drawn from the same distribution, with a $\text{p-value}=0.27$. Overall neither of the groups seems to cluster at the position of the Galactic bar, which is in agreement with the previous studies done by \citet{Dekany2013} and \citet{Minniti2017} as well as seen in the study of the disk RR Lyrae stars by \citep{Dekany2018}. In addition, in the bottom panel of Fig.~\ref{fig:two-point} we do not observe any major trend in RR~Lyrae stars that would imply that they concentrate at the position of the Galactic bar above or below the Galactic plane.

\section{Conclusions} \label{sec:Conclus}

In this paper, we used photometric data for RRab type stars from the OGLE-IV and VVV surveys to study the differences, in physical properties, and spatial distributions of the Oosterhoff I and II populations in the Galactic bulge. Our results can be summarized as follows:

\begin{itemize}
\item We identified and separated the Oo\,I and II populations among non-modulated RR~Lyrae stars in the Galactic bulge, based on the period-amplitude space and using the manifold learning. Approximately 25\,\% of the analyzed stars (from 8\,141 fundamental-mode variables) belong to the Oo\,II group, which is consistent with similar studies for the Galactic halo.

\item Based on the estimated physical parameters the Oo\,II stars are on an average cooler ($\approx$400\,K), more massive ($\approx$0.04\,$M_\odot$), brighter (by 0.23\,mag in $K_{s}$-band), and more metal-poor ($\approx$0.1\,dex)than the Oo\,I pulsators. A comparison with models of stellar pulsation and evolution showed that the Oo\,II variables are more massive than Oo\,I stars.

\item Both Oosterhoff groups were compared with two GCs in the Galactic bulge, NGC~6401 and NGC~6441, identified as containing Oo\,I and Oo\,III stars, respectively. Based on the positions of individual stars in the period-amplitude or metallicity and $\varphi_{21}$ or $\varphi_{31}$ vs. $R_{31}$ diagrams some stars from our Oo\,II group may actually belong to the Oo\,III group. 

\item We studied the spatial distribution of the Oo\,I and II groups in the Galactic bulge. We found that the region denoted in our work as the \textit{foreground} (for distances $d \leq 6.4\,\rm kpc$) of the Galactic bulge contains fewer variables associated with the Oo\,II component (less than 19\,\%, in comparison with the overall distribution 25\,\%). On the other hand, in the central region of the Galactic bulge with one exception, we do not see any difference in the overall value of 25\,\%. We found an increase in the occurrence of Oo\,II stars (above 30\,\%) in the region encompassing the Galactic center in $b$ = (-0.75\,$\pm$0.75 deg) Galactic latitude. The difference in the representation of the Oo\,I and II groups also appears when we progress from the Galactic center toward larger distances to the denoted \textit{background} (for distances $d \geq 10.3\,\rm kpc$) region. Here we found a high occurrence of Oo\,II stars, which constitute almost one-third of the RR~Lyrae stars. Both of these discrepancies may be related to the fact that Oo\,II stars are intrinsically brighter and redder than Oo\,I stars, and therefore detectable at larger distances and in regions with higher extinction.

\item The density profile for both Oosterhoff groups in the Galactic bulge shows that as we move radially outward from the Galactic center the density of the Oo\,II stars rises. In some cases, the normalized density even exceeds the normalized density of the Oo\,I variables with a statistical significance (more than $\sigma$). The distribution of distances from the Galactic bar showed that the Oo\,I group peaks on average on shorter distances than the Oo\,II populations. On the other hand, based on the two-point correlation function for the Oo\,I and II group below the Galactic plane, both groups are similarly distributed. Above the Galactic plane the Oo\,II population seems to be slightly more clustered in comparison with the Oo\,I population. We did not find any correlation between the Galactic bar and the Oosterhoff groups.

\item In order to test our derived extinction law and subsequently our results, we decided to re-do the whole analysis for two independent distance estimations. In the first approach we used the $J$-band from the VVV photometry using the \texttt{pyfiner} \citep{Hajdu2018} for our sample and extinction law derived by \citet{Majaess2016} for $E(J-K)$ to estimate $A_{K}$. In the second approach we used the extinction maps by \citet{Gonzalez2012} to gain $E(J-K)$ and equation 29 from \citet{Nataf2013} to derive $A_{I}$. In the end, we compared the results for all three methods and verified that our main conclusions remain the same regardless of the assumed extinction law.

\end{itemize}

Our main conclusions are as follows. The Galactic bulge contains mainly two Oosterhoff groups (Oo\,I and II) that differ in metallicities, masses, luminosities, and colors. This is in agreement with similar studies in GCs. A small portion of Oo\,II stars probably belong to the Oo\,III group. The spatial distribution of Oo\,I and II groups in the Galactic bulge does not show any systematic difference between both groups and the Oosterhoff populations seem to be well mixed. In the \textit{background} and central region of the Galactic bulge we observe an increase of Oo\,II variables, which is probably an observational bias due to their higher brightness. We do not observe any spatial correlation with the Galactic bar for the aforementioned Oosterhoff groups, which is in agreement with \citet{Dekany2013} and \citet{Minniti2017} but in disagreement with \citet{Pietrukowicz2015}. One of the possible reasons for this discrepancy is a different treatment of extinction, and that our analysis, unlike that of \citeauthor{Pietrukowicz2015}, is partly based on near-infrared data.

\section*{Acknowledgements}

Z.P. acknowledges the support of the Hector Fellow Academy. E.K.G and I.D. were supported by Sonderforschungsbereich SFB 881 "The Milky Way System" (subprojects A2, A3) of the German Research Foundation (DFG). Support for M.C. was provided by Fondecyt through grant \#1171273; the Ministry for the Economy, Development, and Tourism's Millennium Science Initiative through grant IC\,120009, awarded to the Millennium Institute of Astrophysics (MAS); by Proyecto Basal AFB-170002; and by CONICYT's PCI program through grant DPI20140066. R.S. was supported by the National Science Center, Poland, grant agreement DEC-2015/17/B/ST9/03421. M.S. acknowledges support from Postdoc$@$MUNI project CZ.02.2.69/0.0/0.0/16$\_$027/0008360 and GACR international grant 17-01752J.



\newpage 
\bsp	
\label{lastpage}

\begin{thebibliography}{99} 

\bibitem[\protect\citeauthoryear{Alcock et al.}{1997}]{Alcock1997} Alcock C., et al., 1997, ApJ, 474, 217 
\bibitem[\protect\citeauthoryear{Alonso-Garc{\'{\i}}a et al.}{2015}]{Alonso-Garcia2015} Alonso-Garc{\'{\i}}a J., D{\'e}k{\'a}ny I., Catelan M., Contreras Ramos R., Gran F., Amigo P., Leyton P., Minniti D., 2015, AJ, 149, 99 
\bibitem[\protect\citeauthoryear{Armandroff \& Zinn}{1988}]{ArmZin1988} Armandroff T.~E., Zinn R., 1988, AJ, 96, 92 
\bibitem[\protect\citeauthoryear{Asplund et al.}{2009}]{Asplund2009} Asplund M., Grevesse N., Sauval A.~J., Scott P., 2009, ARA\&A, 47, 481 
\bibitem[\protect\citeauthoryear{Bessell, Castelli, \& Plez}{1998}]{Bessell1998} Bessell M.~S., Castelli F., Plez B., 1998, A\&A, 333, 231 
\bibitem[\protect\citeauthoryear{Bla{\v z}ko}{1907}]{Blazhko1907} Bla{\v z}ko S., 1907, AN, 175, 325 
\bibitem[\protect\citeauthoryear{Braga et al.}{2015}]{Braga2015} Braga V.~F., et al., 2015, ApJ, 799, 165 
\bibitem[\protect\citeauthoryear{Cacciari \& Bruzzi}{1993}]{Cacciari1993} Cacciari C., Bruzzi A., 1993, A\&A, 276, 87 
\bibitem[\protect\citeauthoryear{Cacciari, Corwin, \& Carney}{2005}]{Cacciari2005} Cacciari C., Corwin T.~M., Carney B.~W., 2005, AJ, 129, 267 
\bibitem[\protect\citeauthoryear{Carretta et al.}{2009}]{Carretta2009} Carretta E., Bragaglia A., Gratton R., D'Orazi V., Lucatello S., 2009, A\&A, 508, 695 
\bibitem[\protect\citeauthoryear{Catelan}{1992}]{Catelan1992} Catelan M., 1992, A\&A, 261, 457 
\bibitem[\protect\citeauthoryear{Catelan, Pritzl, \& Smith}{2004}]{Catelan2004} Catelan M., Pritzl B.~J., Smith H.~A., 2004, ApJS, 154, 633 
\bibitem[\protect\citeauthoryear{Catelan}{2009}]{Catelan2009} Catelan M., 2009, Ap\&SS, 320, 261 
\bibitem[\protect\citeauthoryear{Clement \& Shelton}{1999}]{Clement1999} Clement C.~M., Shelton I., 1999, ApJ, 515, L85 
\bibitem[\protect\citeauthoryear{Clementini et al.}{2003}]{Clementini2003} Clementini G., Held E.~V., Baldacci L., Rizzi L., 2003, ApJ, 588, L85 
\bibitem[\protect\citeauthoryear{Clementini et al.}{2005}]{Clementini2005} Clementini G., Gratton R.~G., Bragaglia A., Ripepi V., Martinez Fiorenzano A.~F., Held E.~V., Carretta E., 2005, ApJ, 630, L145 
\bibitem[\protect\citeauthoryear{Davis \& Peebles}{1983}]{Davis1983} Davis M., Peebles P.~J.~E., 1983, ApJ, 267, 465 
\bibitem[\protect\citeauthoryear{D{\'e}k{\'a}ny et al.}{2013}]{Dekany2013} D{\'e}k{\'a}ny I., Minniti D., Catelan M., Zoccali M., Saito R.~K., Hempel M., Gonzalez O.~A., 2013, ApJ, 776, L19 
\bibitem[\protect\citeauthoryear{D{\'e}k{\'a}ny et al.}{2018}]{Dekany2018} D{\'e}k{\'a}ny I., Hajdu G., Grebel E.~K., Catelan M., Elorrieta F., Eyheramendy S., Majaess D., Jord{\'a}n A., 2018, ApJ, 857, 54 
\bibitem[\protect\citeauthoryear{Dorfi \& Feuchtinger}{1999}]{Dorfi1999} Dorfi E.~A., Feuchtinger M.~U., 1999, A\&A, 348, 815 
\bibitem[\protect\citeauthoryear{Dotter et al.}{2008}]{Dotter2008} Dotter A., Chaboyer B., Jevremovi{\'c} D., Kostov V., Baron E., Ferguson J.~W., 2008, ApJS, 178, 89-101 
\bibitem[\protect\citeauthoryear{Drake et al.}{2013}]{Drake2013} Drake A.~J., et al., 2013, ApJ, 763, 32 
\bibitem[\protect\citeauthoryear{Fiorentino et al.}{2015}]{Fiorentino2015} Fiorentino G., et al., 2015, ApJ, 798, L12 
\bibitem[\protect\citeauthoryear{Gaia Collaboration et al.}{2018}]{Gaia2018} Gaia Collaboration, et al., 2018, A\&A, 616, A1 
\bibitem[\protect\citeauthoryear{Gonzalez et al.}{2011}]{Gonzalez2011} Gonzalez O.~A., Rejkuba M., Minniti D., Zoccali M., Valenti E., Saito R.~K., 2011, A\&A, 534, L14 
\bibitem[\protect\citeauthoryear{Gonzalez et al.}{2012}]{Gonzalez2012} Gonzalez O.~A., Rejkuba M., Zoccali M., Valenti E., Minniti D., Schultheis M., Tobar R., Chen B., 2012, A\&A, 543, A13 
\bibitem[\protect\citeauthoryear{Grevesse \& Sauval}{1998}]{Grevesse1998} Grevesse N., Sauval A.~J., 1998, SSRv, 85, 161 
\bibitem[\protect\citeauthoryear{Hajdu et al.}{2015}]{Hajdu2015} Hajdu G., Catelan M., Jurcsik J., D{\'e}k{\'a}ny I., Drake A.~J., Marquette J.-B., 2015, MNRAS, 449, L113 
\bibitem[\protect\citeauthoryear{Hajdu et al.}{2018}]{Hajdu2018} Hajdu G., D{\'e}k{\'a}ny I., Catelan M., Grebel E.~K., Jurcsik J., 2018, ApJ, 857, 55 
\bibitem[\protect\citeauthoryear{Hamilton}{1993}]{Hamilton1993} Hamilton A.~J.~S., 1993, ApJ, 417, 19 
\bibitem[\protect\citeauthoryear{Harris}{1996}]{Harris1996} Harris W.~E., 1996, AJ, 112, 1487 
\bibitem[\protect\citeauthoryear{Haschke, Grebel, \& Duffau}{2012}]{Haschke-2012-SMC-LMC} Haschke R., Grebel E.~K., Duffau S., 2012, AJ, 144, 106
\bibitem[\protect\citeauthoryear{Howard et al.}{2009}]{Howard2009} Howard C.~D., et al., 2009, ApJ, 702, L153 
\bibitem[\protect\citeauthoryear{Iglesias \& Rogers}{1996}]{Iglesias1996} Iglesias C.~A., Rogers F.~J., 1996, ApJ, 464, 943 
\bibitem[\protect\citeauthoryear{Ivezi{\'c} et al.}{2014}]{Ivezic2014} Ivezi{\'c} {\v Z}., Connelly A.~J., VanderPlas J.~T., Gray A., 2014, Statistics, Data Mining, and Machine Learning in Astronomy, Princeton Univ. Press, Princeton, NJ
\bibitem[\protect\citeauthoryear{Jacyszyn-Dobrzeniecka et al.}{2017}]{JD2massive016RRLyr} Jacyszyn-Dobrzeniecka A.~M., et al., 2017, AcA, 67, 1 
\bibitem[\protect\citeauthoryear{Johnson}{1965}]{Johnson1965} Johnson H.~L., 1965, ApJ, 141, 923 
\bibitem[\protect\citeauthoryear{Jurcsik}{1995}]{Jurcsik1995} Jurcsik J., 1995, AcA, 45, 653  
\bibitem[\protect\citeauthoryear{Jurcsik \& Kov\'acs}{1996}]{JK1996} Jurcsik J., Kov\'acs G., 1996, A\&A, 312, 111 
\bibitem[\protect\citeauthoryear{Jurcsik et al.}{2018}]{Jurcsik2018} Jurcsik J., Hajdu G., D{\'e}k{\'a}ny I., Nuspl J., Catelan M., Grebel E.~K., 2018, MNRAS, 475, 4208 
\bibitem[\protect\citeauthoryear{Kormendy \& Kennicutt}{2004}]{Kormendy2004} Kormendy J., Kennicutt R.~C., Jr., 2004, ARA\&A, 42, 603 
\bibitem[\protect\citeauthoryear{Kunder \& Chaboyer}{2009}]{Kunder2009} Kunder A., Chaboyer B., 2009, AJ, 138, 1284 
\bibitem[\protect\citeauthoryear{Kunder et al.}{2016}]{Kunder2016} Kunder A., et al., 2016, ApJ, 821, L25
\bibitem[\protect\citeauthoryear{Kunder et al.}{2018}]{Kunder2018} Kunder A., et al., 2018, AJ, 155, 171 
\bibitem[\protect\citeauthoryear{Kurucz}{2005}]{Kurucz2005} Kurucz R.~L., 2005, Mem. Soc. Astron. Ital. Suppl., 8, 14
\bibitem[\protect\citeauthoryear{Landy \& Szalay}{1993}]{LS1993} Landy S.~D., Szalay A.~S., 1993, ApJ, 412, 64 
\bibitem[\protect\citeauthoryear{Li{\v s}ka et al.}{2016}]{Liska2016} Li{\v s}ka J., Skarka M., Zejda M., Mikul{\'a}{\v s}ek Z., de Villiers S.~N., 2016, MNRAS, 459, 4360 
\bibitem[\protect\citeauthoryear{L{\"u}tticke, Dettmar, \& Pohlen}{2000}]{Lutticke2000} L{\"u}tticke R., Dettmar R.-J., Pohlen M., 2000, A\&AS, 145, 405 
\bibitem[\protect\citeauthoryear{Majaess et al.}{2016}]{Majaess2016} Majaess D., Turner D., D{\'e}k{\'a}ny I., Minniti D., Gieren W., 2016, A\&A, 593, A124 
\bibitem[\protect\citeauthoryear{Marconi et al.}{2015}]{Marconi2015} Marconi M., et al., 2015, ApJ, 808, 50 
\bibitem[\protect\citeauthoryear{McNamara \& Barnes}{2014}]{McNamara2014} McNamara D.~H., Barnes J., 2014, AJ, 147, 31 
\bibitem[\protect\citeauthoryear{McWilliam \& Zoccali}{2010}]{McWilliam2010} McWilliam A., Zoccali M., 2010, ApJ, 724, 1491 
\bibitem[\protect\citeauthoryear{Miceli et al.}{2008}]{Miceli2008} Miceli A., et al., 2008, ApJ, 678, 865 
\bibitem[\protect\citeauthoryear{Minniti et al.}{2010}]{Minniti2010} Minniti D., et al., 2010, NewA, 15, 433 
\bibitem[\protect\citeauthoryear{Minniti et al.}{2017}]{Minniti2017} Minniti D., et al., 2017, AJ, 153, 179 
\bibitem[\protect\citeauthoryear{Monelli et al.}{2017}]{Monelli2017} Monelli M., Fiorentino G., Bernard E.~J., Mart{\'{\i}}nez-V{\'a}zquez C.~E., Bono G., Gallart C., Dall'Ora M., Stetson P.~B., 2017, ApJ, 842, 60 
\bibitem[\protect\citeauthoryear{Muraveva et al.}{2015}]{Muraveva2015} Muraveva T., et al., 2015, ApJ, 807, 127 
\bibitem[\protect\citeauthoryear{Muraveva et al.}{2018}]{Muraveva2018} Muraveva T., Delgado H.~E., Clementini G., Sarro L.~M., Garofalo A., 2018, MNRAS, 481, 1195 
\bibitem[\protect\citeauthoryear{Nataf et al.}{2013}]{Nataf2013} Nataf D.~M., et al., 2013, ApJ, 769, 88 
\bibitem[\protect\citeauthoryear{Neeley et al.}{2015}]{Neeley2015} Neeley J.~R., et al., 2015, ApJ, 808, 11 
\bibitem[\protect\citeauthoryear{Ness et al.}{2012}]{Ness2012} Ness M., et al., 2012, ApJ, 756, 22 
\bibitem[\protect\citeauthoryear{Nishiyama et al.}{2005}]{Nishiyama2005} Nishiyama S., et al., 2005, ApJ, 621, L105 
\bibitem[\protect\citeauthoryear{Oosterhoff}{1939}]{Oosterhoff1939} Oosterhoff P.~T., 1939, Obs, 62, 104
\bibitem[\protect\citeauthoryear{Oosterhoff}{1944}]{Oosterhoff1944} Oosterhoff P.~T., 1944, BAN, 10, 55 
\bibitem[\protect\citeauthoryear{Papadakis et al.}{2000}]{Papadakis2000} Papadakis I., Hatzidimitriou D., Croke B.~F.~W., Papamastorakis I., 2000, AJ, 119, 851  
\bibitem[\protect\citeauthoryear{Peebles}{1980}]{Peebles1980} Peebles P.~J.~E., 1980, NYASA, 336, 161 
\bibitem[\protect\citeauthoryear{Pedregosa et al.}{2011}]{Pedregosa2011} Pedregosa F., Varoquaux, G., Gramfort, A., et al. 2011, Journal of Machine Learning Research, 12, 2825
\bibitem[\protect\citeauthoryear{Pietrinferni et al.}{2004}]{Pietrinferni2004} Pietrinferni A., Cassisi S., Salaris M., Castelli F., 2004, ApJ, 612, 168 
\bibitem[\protect\citeauthoryear{Pietrukowicz et al.}{2015}]{Pietrukowicz2015} Pietrukowicz P., et al., 2015, ApJ, 811, 113 
\bibitem[\protect\citeauthoryear{Pritzl et al.}{2000}]{Pritzl2000} Pritzl B., Smith H.~A., Catelan M., Sweigart A.~V., 2000, ApJ, 530, L41 
\bibitem[\protect\citeauthoryear{Prudil et al.}{2017}]{PrudilSmolec2017} Prudil Z., Smolec R., Skarka M., Netzel H., 2017, MNRAS, 465, 4074
\bibitem[\protect\citeauthoryear{Prudil \& Skarka}{2017}]{Prudil2017} Prudil Z., Skarka M., 2017, MNRAS, 466, 2602 
\bibitem[\protect\citeauthoryear{Prudil et al.}{2018}]{PrudilOO2018} Prudil Z., Grebel E.~K., D{\'e}k{\'a}ny I., Smolec R., Skarka M., 2018, in The RR Lyrae 2017 Conference. Revival of the Classical Pulsators: from Galactic Structure to Stellar Interior Diagnostics, ed. R. Smolec, K. Kinemuchi, \& R. I. Anderson, Vol. 6, 37-41
\bibitem[\protect\citeauthoryear{Sandage, Katem, \& Sandage}{1981}]{Sandage-K-S-1981} Sandage A., Katem B., Sandage M., 1981, ApJS, 46, 41 
\bibitem[\protect\citeauthoryear{Sandage}{1981}]{Sandage1981} Sandage A., 1981, ApJ, 248, 161
\bibitem[\protect\citeauthoryear{Sandage}{2004}]{Sandage2004} Sandage A., 2004, AJ, 128, 858 
\bibitem[\protect\citeauthoryear{Sandage}{2006}]{Sandage2006} Sandage A., 2006, AJ, 131, 1750 
\bibitem[\protect\citeauthoryear{Sesar et al.}{2013}]{Sesar2013} Sesar B., et al., 2013, AJ, 146, 21 
\bibitem[\protect\citeauthoryear{Smith}{1995}]{Smith1995} Smith H.~A., 1995, CAS, 27, 
\bibitem[\protect\citeauthoryear{Smolec}{2005}]{Smolec2005} Smolec R., 2005, AcA, 55, 59 
\bibitem[\protect\citeauthoryear{Smolec \& Moskalik}{2008}]{Smolec2008} Smolec R., Moskalik P., 2008, AcA, 58, 193 
\bibitem[\protect\citeauthoryear{Smolec}{2016}]{Smolec2016BL} Smolec R., 2016, Proceedings of the Polish Astron. Soc., 3, 22 
\bibitem[\protect\citeauthoryear{Smolec et al.}{2016}]{Smolec2016} Smolec R., Prudil Z., Skarka M., Bakowska K., 2016, MNRAS, 461, 2934 
\bibitem[\protect\citeauthoryear{Soszy{\'n}ski et al.}{2014}]{Soszynski2014} Soszy{\'n}ski I., et al., 2014, AcA, 64, 177 
\bibitem[\protect\citeauthoryear{Soszy{\'n}ski et al.}{2017}]{Soszynski2017} Soszy{\'n}ski I., et al., 2017, AcA, 67, 297 
\bibitem[\protect\citeauthoryear{Szab{\'o}}{2014}]{Szabo2014} Szab{\'o} R., 2014, IAUS, 301, 241 
\bibitem[\protect\citeauthoryear{Udalski et al.}{1992}]{Udalski1992} Udalski A., Szymanski M., Kaluzny J., Kubiak M., Mateo M., 1992, AcA, 42, 253 
\bibitem[\protect\citeauthoryear{Udalski, Szyma{\'n}ski, \& Szyma{\'n}ski}{2015}]{Udalski2015} Udalski A., Szyma{\'n}ski M.~K., Szyma{\'n}ski G., 2015, AcA, 65, 1
\bibitem[\protect\citeauthoryear{Tenenbaum, de Silva, \& Langford}{2000}]{Tenenbaum2000} Tenenbaum J.~B., de Silva V., Langford J.~C., 2000, Sci, 290, 2319 
\bibitem[\protect\citeauthoryear{Tsapras et al.}{2017}]{Tsapras2017} Tsapras Y., et al., 2017, MNRAS, 465, 2489 
\bibitem[\protect\citeauthoryear{van Albada \& Baker}{1973}]{vanAlbada1973} van Albada T.~S., Baker N., 1973, ApJ, 185, 477 
\bibitem[\protect\citeauthoryear{Wegg \& Gerhard}{2013}]{Wegg2013} Wegg C., Gerhard O., 2013, MNRAS, 435, 1874 
\bibitem[\protect\citeauthoryear{Zinn \& West}{1984}]{Zinn1984} Zinn R., West M.~J., 1984, ApJS, 55, 45 
\end{thebibliography}
\end{document}